%% file: 0-main-tpds.tex
\documentclass[manuscript,screen]{acmart}

\AtBeginDocument{%
  \providecommand\BibTeX{{%
    \normalfont B\kern-0.5em{\scshape i\kern-0.25em b}\kern-0.8em\TeX}}}

\usepackage{xcolor,colortbl}
\usepackage{tabularx}
\usepackage{multirow}
\usepackage{color}
\usepackage{listings}
\usepackage{lstautogobble}
\usepackage{xspace}
\usepackage{amsmath}
\usepackage{hyperref}
\usepackage{soul}

\makeatletter
\AtBeginDocument{%
  \let\c@figure\c@lstlisting
  
  \let\ftype@lstlisting\ftype@figure 
}
\makeatother

\usepackage{subcaption}
\usepackage{caption}

\definecolor{light}{rgb}{0.5, 0.5, 0.5}
\def\light#1{{\color{light}#1}}
\newcommand{\keywordSP}[1]{{\color{blue}{#1}}}

\lstdefinelanguage{NEAR}
{
  morekeywords= [1]{ COPY,
  ENTRYPOINT, function, fixedPoint, filter, Min, forall},
  morekeywords = [2]{propNode, Graph, node, edge, int, bool, in, until, for, if, from, do, while, void, __global__, unsigned},
  morekeywords = [3]{attachNodeProperty, nodes, neighbors, get_edge},
  morekeywords = [4]{True, False},
  morecomment=[l]{\//},
  morestring=[b]",
}

\lstdefinestyle{nolinenum}{
numbers=none
}
\lstdefinestyle{mystyle}{
  xleftmargin=.1\textwidth, 
  numbers=left, 
  numbersep=8pt,                  
  backgroundcolor=\color{white},  
  showspaces=false,
  rulecolor=\color{black},
  stringstyle=\color{green},
  numberstyle=\scriptsize\color{gray},
  commentstyle=\color{gray},
  identifierstyle=\color{black},
  keywordstyle = [1]\color{purple},
  keywordstyle=[2]\color{blue},
  keywordstyle=[3]\color{black},
  keywordstyle=[4]\color{green},
  keywordstyle=\color{purple}\bfseries,
  showstringspaces=false,        
  showtabs=false,                
  tabsize=1,                      
  captionpos=b,                   
  breaklines=true,                
  breakatwhitespace=true,         
  title=\lstname , 
}
\lstdefinestyle{cstyle}{
 xleftmargin=.1\textwidth, 
  numbers=left,
  numbersep=8pt,                  
  backgroundcolor=\color{white},  
  showspaces=false,
  rulecolor=\color{black},
  stringstyle=\color{green},
  numberstyle=\scriptsize\color{gray},
  commentstyle=\color{gray},
  identifierstyle=\color{black},
  keywordstyle=\color{purple},
  showstringspaces=false,        
   showtabs=false,                
  tabsize=1,                      
  captionpos=b,                   
  breaklines=true,                
  breakatwhitespace=true,         
  title=\lstname 
}

\begin{document}

\title{\name: A Versatile DSL for Graph Analytics}

\author{Nibedita Behera }
\email{cs20s023@cse.iitm.ac.in}
\orcid{0000-0002-1563-8686}

\author{Ashwina Kumar}
\email{cs20d016@cse.iitm.ac.in}
\orcid{0000-0001-6425-7479}

\author{Ebenezer Rajadurai T}
\email{ebenezerrajadurai5@gmail.com}

\author{Sai Nitish}
\email{bsainitishkumar@gmail.com}

\author{Rajesh Pandian M}
\email{mrprajesh@cse.iitm.ac.in}
\orcid{0000-0003-4702-4678}

\author{Rupesh Nasre}
\email{rupesh@cse.iitm.ac.in}
\orcid{0000-0001-7490-625X}

\affiliation{%
  \institution{IIT Madras}
  \country{India.}
}
\renewcommand{\shortauthors}{Behera, et al.}

\newcommand{\name}{StarPlat\xspace} 

\newcommand{\bc}{betweenness centrality\xspace} 
\newcommand{\pr}{page rank\xspace} 
\newcommand{\sssp}{single-source shortest paths\xspace} 
\newcommand{\tc}{triangle counting\xspace} 
\newcommand{\bcc}{Betweenness Centrality\xspace} 
\newcommand{\prr}{PageRank\xspace} 
\newcommand{\ssspp}{Single-Source Shortest Paths\xspace} 
\newcommand{\tcc}{Triangle Counting\xspace} 

\newcommand{\gr}{Gunrock\xspace} 
\newcommand{\lsgpu}{LonestarGPU\xspace} 

\newcommand{\etodo}[1]{{\color{red}{#1}}}
\newcommand{\speedup}{ABC $\times$\xspace} 
\newcommand{\rtodo}[1]{{\color{blue}{rajz: #1}.}}
\newcommand{\tocite}{{\color{blue}{CITE}}\xspace}
\newcommand{\todo}[1]{{\color{blue}{#1}}}
\newcommand{\mypara}[1]{\vspace{1mm}\noindent\textbf{#1.} }
\newcommand{\REM}[1]{}

\input{0ABS.tex}

\ccsdesc[500]{Computing methodologies~Massively parallel algorithms}
\ccsdesc[300]{Computing methodologies~Shared memory algorithms}
\ccsdesc[500]{Computing methodologies~Parallel programming languages}
\ccsdesc[500]{Software and its engineering~Parallel programming languages}

\keywords{Graph Algorithms, Domain-Specific Language, OpenMP, MPI, CUDA}

\maketitle

\input{1INT.tex}

\input{2AST.tex}

\input{3GEN.tex}

\input{4OPT.tex}
\input{5EXP.tex}
\input{6REL.tex}
\input{7CON.tex}

\begin{acks}
We gratefully acknowledge the use of the computing resources at HPCE, IIT Madras. This work is supported by India's National Supercomputing Mission grant CS1920/1123/MEIT/008606.
\end{acks}

\bibliographystyle{acmart}
\input{0-main-tpds.bbl}

\appendix
\input{91codes}

\end{document}

%% file: 0ABS.tex
\begin{abstract}
Graphs model several real-world phenomena. With the growth of unstructured and semi-structured data, parallelization of graph algorithms is inevitable. Unfortunately, due to inherent irregularity of computation, memory access, and communication, graph algorithms are traditionally challenging to parallelize. To tame this challenge, several libraries, frameworks, and domain-specific languages (DSLs) have been proposed to reduce the parallel programming burden of the users, who are often domain experts. However, existing frameworks to model graph algorithms typically target a single architecture.
In this paper, we present a graph DSL, named \name, that allows programmers to specify graph algorithms in a high-level format, but generates code for three different backends from the same algorithmic specification. In particular, the DSL compiler generates OpenMP for multi-core systems, MPI for distributed systems, and CUDA for many-core GPUs. Since these three are completely different parallel programming paradigms, binding them together under the same language is challenging. We share our experience with the language design. Central to our compiler is an intermediate representation which allows a common representation of the high-level program, from which individual backend code generations begin. We demonstrate the expressiveness of \name~by specifying four graph algorithms: betweenness centrality computation, page rank computation, single-source shortest paths, and triangle counting. Using a suite of ten large graphs, we illustrate the effectiveness of our approach by comparing the performance of the generated codes with that obtained with hand-crafted library codes. We find that the generated code is competitive to library-based codes in many cases. More importantly, we show the feasibility to generate efficient codes for different target architectures from the same algorithmic specification of graph algorithms.

\end{abstract}

%% file: 1INT.tex
\section{Introduction} \label{sec intro}

The graph data structure has become an integral component of many real-world applications for modelling relationships in their data today. Enormous growth of unstructured and semi-structured data has led to these graphs growing to billions of edges. Therefore, parallel graph analytic solutions are inevitable to scale to such large graph sizes. The last two decades have witnessed significant advances in hardware towards parallel processing, which has also resulted in major developments in the software support, be it in the form of libraries or programming languages. These hardware and software architectures are primarily suited for regular codes wherein the data access, control flow, and communication patterns are statically identifiable. As an example, tiling of regular matrix computations can now be performed automatically by the compiler for improved cache benefits or with minimal communication.

Unfortunately, existing widely-used compilers perform poorly in the case of graph algorithms. This is due to the inherent \textit{irregularity} in sparse graph processing, wherein the access patterns are dependent on the input (which is unavailable at compile time). On the other hand, our community has shown that graph algorithms exhibit enough parallelism to keep the cores busy on several real-world graphs~\cite{kulkarni-pldi12}. Unfortunately, manually exploiting this parallelism on various hardware is quite challenging. The programmer needs to be an expert in the application domain to exploit algorithmic properties, in high-performance computing (HPC) to model the computation to suit the target hardware, and also in computing systems to optimize the overall application and the support software on the given infrastructure. Even if corporations and institutions can find and afford such an expert, the expert is unlikely to be a best-fit for another application domain. A sore practical reality is that we have domain experts who are not HPC or systems experts, and we have HPC experts who may not know enough about the underlying application domain.

A viable alternative towards improved productivity is a graph library or a domain-specific language, which allows domain experts to express their algorithm using API or high-level constructs, and the library or the DSL compiler taking care of generating high-performing code for the target hardware. A range of such frameworks exists today~\cite{Galois, Gunrock, GreenMarl, GraphIt}. 
Multiple of these frameworks partially or fully hide the parallelism intricacies, provide mnemonics for scheduling strategies, and perform program analysis to identify races to generate synchronization code. It is often seen that the amount of code one needs to write reduces considerably compared to that in a hand-crafted explicitly parallel code. 

One of the limitations of the existing frameworks (libraries or graph DSLs) is that they target a specific hardware architecture. For instance, 
Ligra~\cite{Ligra} is a graph processing frameworks specifically for shared memory systems. Greenmarl~\cite{GreenMarl} primarily targets multicore devices and supports distributed implementation through Pregel API. Gunrock~\cite{Gunrock} is a CUDA library for graph processing on GPUs. GraphIt~\cite{GraphIt} is a DSL that targets shared memory and manycore systems. Rarely, a framework targets more than one backends (e.g., Kokkos~\cite{kokkos3-2022}). For the best performance and considering the range of HPC hardware we use today, libraries often restrict themselves to a certain target. Multi-core parallelism permits data shared across all the running threads, while a cluster-level parallelism demands explicit communication, while further, many-core parallelism necessitates a hierarchical computation and SIMD execution for the best performance. While domain-experts are aware of these basic differences, they should be able to \textit{specify} rather than \textit{code up} these for an architecture to achieve desired performance. Therefore, it is ideal if the domain-specific language encompasses different backend peculiarities in its design. This bears the advantage of simplified and efficient code generation, and avoids \textit{patching} a language originally designed for a different parallel architecture.

In this work, we propose a graph DSL named \name which allows a user to provide an algorithm specification of graph problems using its high-level graph specific constructs and generates code for multiple backends from the same algorithmic specification (currently, multi-core, distributed, and many-core).
The constructs are carefully designed to abstract the  parallelization specific implementations from users, while ensuring generation of  high performing graph codes. The \name compiler moulds the high level information embedded in these constructs to different  architecture specific implementations. 
The compiler translates the DSL code to C/C++. It uses OpenMP parallelism for the multicore setting, MPI for the distributed setup, and CUDA for the many-core architecture. 
Encompassing these three very different backends in the same DSL compiler is challenging, and we highlight these challenges in this manuscript. We proudly admit that we build upon the insights provided by the existing libraries and DSLs, borrow certain constructs from these frameworks (while providing our own), and generate code competitive to these frameworks in terms of performance.
In particular, we illustrate the versatility of \name with a discussion on four graph algorithms: Betweenness Centrality~(BC), PageRank~(PR), Single Source Shortest Paths~(SSSP), and Triangle Counting~(TC).

\begin{figure}
    \centering
    \includegraphics[scale=0.6]{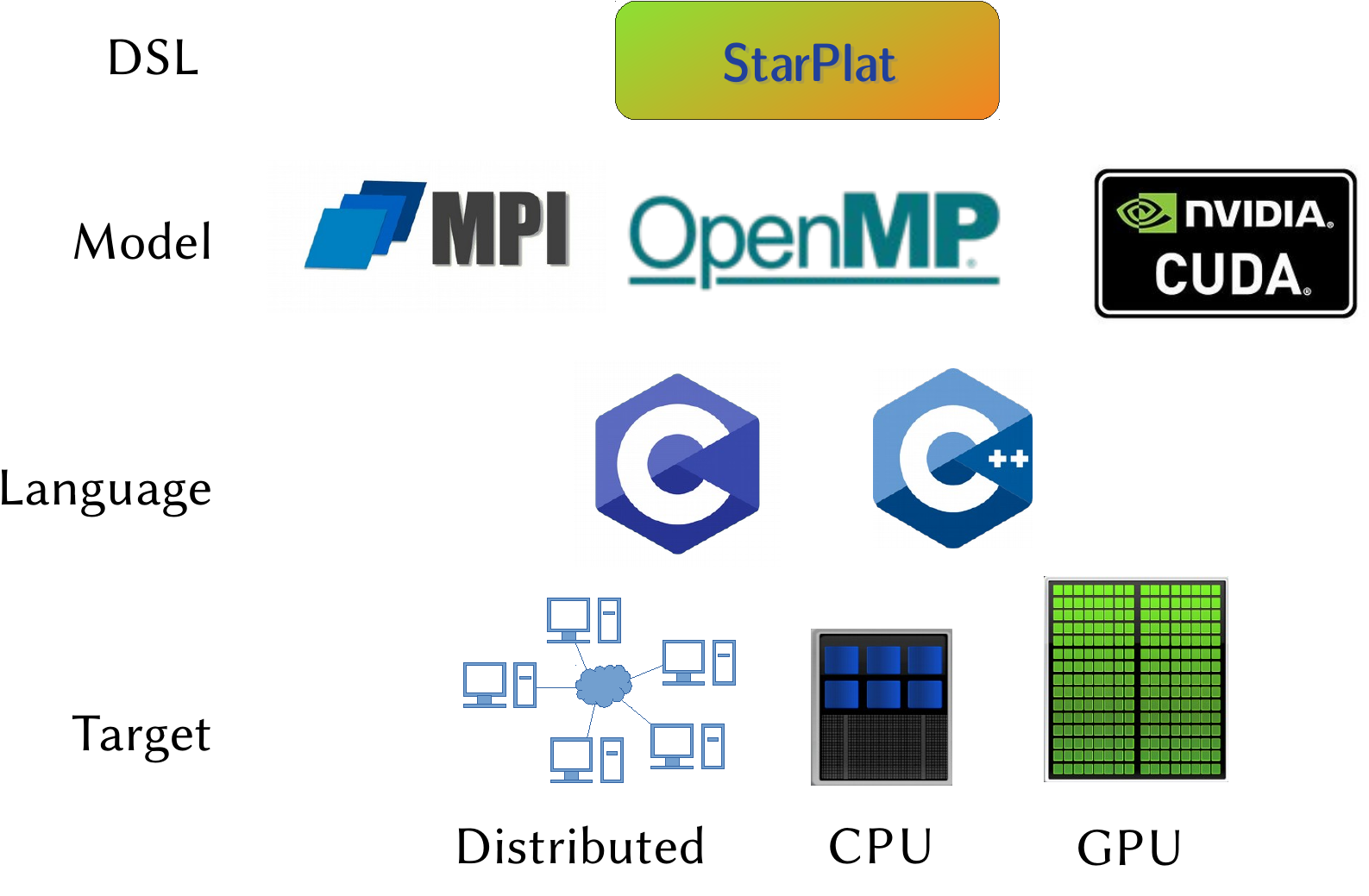}
    \caption{Overview of \name compiler 
    }
    \label{fig: overview}
\end{figure}

This work makes the following technical contributions:
\begin{itemize}
    \item \name\footnote{ \url{https://github.com/nibeditabh/StarPlat} 
    }, a DSL for graph analytics which allows users to write a high level algorithmic specification of their static graph processing, which captures the parallelism intentions but decouples the target architecture.
    \item An intermediate representation common across backends that captures all the essential information associated with constructs for their translation in the target architecture.
    \item A code-generation scheme to support translation of the intermediate representation into efficient codes for the multicore (using OpenMP), distributed (using OpenMPI), and many-core (using CUDA) backends.
    \item Performance analysis of the generated BC, PR, SSSP, and TC codes for different target hardware on a variety of popular graphs, and illustrating competitive performance against that of the hand-crafted frameworks.
\end{itemize}

The rest of the article is organized as follows: Section~\ref{sec language and ast} presents the language specification and the intermediate representation. Section~\ref{sec code gen} describes the code-generation scheme followed for the translation of the DSL code for each backend.
Section~\ref{sec optimizations} provides an overview of the backend-specific optimizations \name employs for efficient code generation. The experimental evaluation of the generated code for each backend is discussed in Section~\ref{sec exp evaluation}. Section~\ref{sec relatedwork} discusses the related work for graph analytics. We summarise our experience and conclude in Section~\ref{sec conclusion}.

%% file: 2AST.tex
\section{\name Language and Frontend} \label{sec language and ast}
The high-level philosophy of \name is to relieve the user of the parallelization constructs as much as possible, and to achieve performance competitive to hand-tuned codes by providing constructs and hints on aggregates. In rare cases, when it is a must to have a trade-off between abstraction and performance, we have taken a conscious decision to prioritize abstraction. This is because the language is meant primarily for domain-experts (rather than HPC experts).

From day one, the language was designed to abstract away the hardware. This was a challenge, since the backends are quite different. But we have found commonalities at the algorithmic level which we encode using specific constructs, which could be then translated to the appropriate backend code. For instance, when a lock-based synchronization was required in OpenMP, it also demanded communication in MPI, and a lock-free synchronization in CUDA.

\name provides various abstractions and data types relevant to static graph algorithm, such as \texttt{Graph}, \texttt{node}, \texttt{edge}, \texttt{propNode} (for node property) etc. The programmer writes in a procedural style and hints at the parallelization opportunities using aggregate iteration constructs such as \texttt{forall} along with other inherently parallel operations. The compiler decides whether to exploit this parallelization (e.g., nested \texttt{forall}).
Several algorithms can be viewed as iterative procedures repeatedly executed until convergence. The convergence criteria is application dependent. Such an iterative procedure is described using a \texttt{fixedPoint} construct. \name follows a "batteries included" approach and has several utility functions for the data types. Internally, each of the functions is implemented target-hardware-wise. 

\subsection{Compiler Overview}
\begin{figure}[H]
\centering
\includegraphics[scale=0.6]{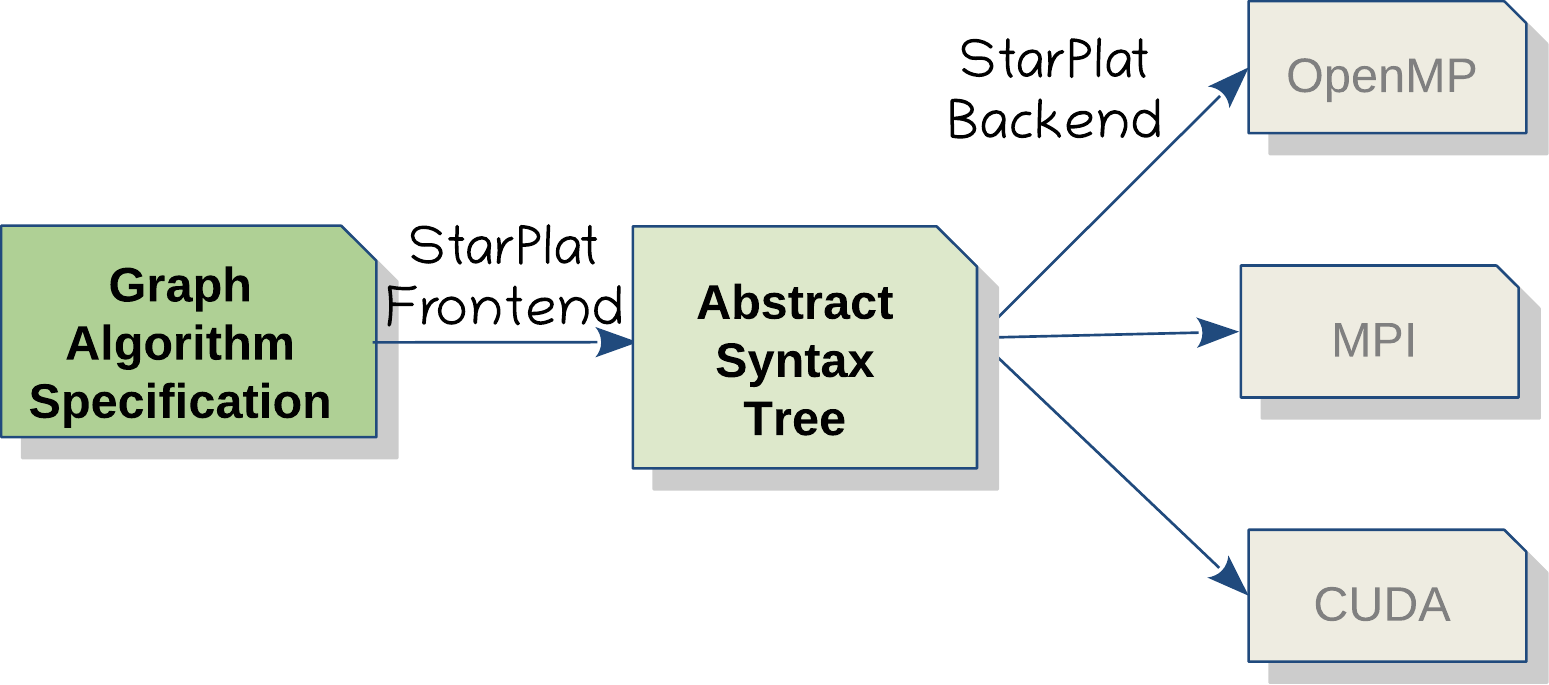}
\caption{Process flow of the \name compiler} 
\label{fig:compiler}
\end{figure}

As in a general-purpose compiler, \name's compiler is split into frontend and backend, as shown in Figure~\ref{fig:compiler}. 
The frontend is responsible for ensuring syntactic and semantic consistency of the \name implementation. The frontend builds an abstract syntax tree (AST) of the supplied input code. 
The AST is common across all the backends and is populated with the metadata for each construct during the parsing stage of the compiler. It is then fed to the appropriate code generator depending upon the target code the user wishes to generate.

\subsection{Language Overview with an Example: SSSP}
Single source shortest path computation~(SSSP) finds a shortest path from a designated source node to every other node in the graph. Parallel SSSP computation can be modelled as an iterative procedure, as illustrated in Figure~\ref{sssp-dsl-sample}. This is a variant of the Bellman-Ford's algorithm.

\begin{lstlisting}[language=NEAR, style=mystyle, label=sssp-dsl-sample, caption = SSSP computation in \name %~\rtodo{ RN, Is running Fig+Code numbers  okay? Codes named as Figs is ok?}
, xleftmargin=.1\textwidth
]
function computeSSSP (Graph g, node src) {
    propNode<int> dist; /*@\label{computeSSSP-stat-3}*/   
    propNode<bool> modified;  /*@\label{computeSSSP-stat-4}*/
    g.attachNodeProperty(dist = INF, modified = False);  /*@\label{computeSSSP-stat-5}*/
    src.modified = True;  /*@\label{computeSSSP-stat-6}*/
    src.dist=0;  /*@\label{computeSSSP-stat-7}*/
    bool finished = False;  /*@\label{computeSSSP-stat-8}*/
    fixedPoint until (finished: !modified) {/*@\label{computeSSSP-stat-9}*/    
        forall (v in g.nodes().filter(modified == True))  { /*@\label{computeSSSP-stat-11}*/ 
            forall (nbr in g.neighbors(v)) { /*@\label{computeSSSP-stat-13}*/           
                edge e = g.getEdge(v, nbr);  /*@\label{computeSSSP-stat-15}*/
                <nbr.dist, nbr.modified> = <Min(nbr.dist, v.dist + e.weight), True>; /*@\label{computeSSSP-stat-16}*/ 
}   }   }   }
\end{lstlisting}

The function \texttt{compute\_SSSP} takes two arguments: a graph \light{g} and source vertex \light{src}. Statement~\ref{computeSSSP-stat-3} declares a variable \light{dist} of type \texttt{propNode}. \texttt{propNode\textless int\textgreater} marks \light{dist} as an integer type attribute associated with each vertex.
Statement~\ref{computeSSSP-stat-4} creates a \texttt{bool} type vertex attribute \light{modified}. In Statement~\ref{computeSSSP-stat-5}, the function \texttt{attachNodeProperty} associates the two attributes with each vertex of \light{g}, and initializes the property values of \light{dist} to infinity (\texttt{INT\_MAX}) and modified to false, as aggregate operations.
In Statements~\ref{computeSSSP-stat-6}--\ref{computeSSSP-stat-7}, the \light{dist} and \light{modified} attributes for vertex \light{src} are assigned specific values. In particular, the distance of \light{src} is set to zero, and the vertex is also marked as modified. The \texttt{fixedPoint} construct in Statement~\ref{computeSSSP-stat-9} makes sure the iterative procedure executes until the variable \light{finished} becomes true. The variable \light{finished} is updated in each iteration, and decides if another iteration is necessary. A fixed-point is reached when no vertices are modified. This is achieved by or-ing the \light{modified} attribute of each node and assigning to \light{finished}.
The iteration construct \light{forall} in Statement~\ref{computeSSSP-stat-11} specifies parallel iteration over the graph nodes. In addition, the optional \light{filter} clause allows processing a subset of the vertices meeting a criterion.
The \light{forall} construct in Statement~\ref{computeSSSP-stat-13} specifies iterating over all the neighbors of vertex \texttt{v}  in parallel and for each neighbor \light{nbr}, the corresponding edge is relaxed (Statements~\ref{computeSSSP-stat-15}--\ref{computeSSSP-stat-16}). 
The multiple assignment construct in Statement~\ref{computeSSSP-stat-16} performs multiple tasks and is a suitable example of the abstraction provided by the DSL's constructs (borrowed from Green-Marl~\cite{GreenMarl}). The rvalue  \texttt{Min(nbr.dist, v.dist + e.weight)} is assigned to the \light{dist} attribute of \texttt{nbr} with the sum of \texttt{v.dist} and \texttt{e.weight} based on  whether the alternative distance via \light{v} is smaller than the existing distance of \texttt{nbr}. If \texttt{nbr.dist} is updated, the \light{modified} attribute of \texttt{nbr} is set. Such a construct allows processing related statements together (as a critical section, to be precise) and translates to an update logic protected by some form of synchronization to prevent data races.

We highlight that the SSSP code structurally resembles the algorithm specified in a textbook, with a few changes.
Hence, \name eases implementation, analysis, and modification of graph algorithms from the context of parallelization. The multi-core translation of this code would follow a similar structure as the \name code with OpenMP pragmas inserted; the distributed version differs in terms of complexities involved in scattering and gathering data across devices; while a many-core version need to pre-transfer the data being used inside \texttt{forall} from CPU to GPU.

We now delve deeper into various \name constructs.

\subsection{Language Constructs}
We first discuss various data types, followed by iteration schemes and reductions.

\subsubsection{Data Types}

\name supports graph theoretic concepts as first-class types such as Graph, node, edge, node attribute, edge attribute, etc. It also supports primitive data types: \texttt{int}, \texttt{bool}, \texttt{long}, \texttt{float}, and \texttt{double}.

The \texttt{Graph} data type encapsulates the operations and properties of a standalone graph. The properties include its nodes, edges, number of nodes, number of edges, etc. 
\name stores the graph in Compressed Sparse Row (CSR) format, which provides the storage benefits of adjancency lists, and also allows seamless transfer across devices, due to the use of offsets.
The data type also facilitates information gathering and manipulation at the node and the edge levels. Since the nodes and edges are tightly bound to the graph, it becomes convenient for Graph to support this through various library functions. For instance, as per the semantics of \texttt{neighbors(u)}, it returns the outgoing neighbors in case of a directed graph and all the neighbors in case of an undirected one. For directed graphs, graph type also exposes a function that returns the incoming neighbours to a node, \texttt{nodesTo()}. This needs Graph to maintains a CSR representation for also its transpose, which becomes handy in algorithms which perform computation on a transposed version of the input graph (e.g., Betweenness Centrality).


\begin{lstlisting}[language=NEAR, style=mystyle, label=toy-code,  caption = Example program to illustrate various data types in \name %\todo{Let's show some useful toy example.} %\rtodo{could not get center align for this code.-DONE}, 
]
function foo(Graph g, propNode<int> modified, propNode<int> data) {
    SetN<g> nodeSet; /*@\label{toy-stat-2}*/
    for(v in g.nodes().filter(modified)) { // Sequential loop /*@\label{toy-stat-3}*/
        for(nbr in g.neighbors(v)) {  /*@\label{toy-stat-4}*/
            if(nbr.data > THRESHOLD) {
                nodeSet.addNode(nbr); /*@\label{toy-stat-8}*/ 	  	     
}   }   }   }
\end{lstlisting}    

A node and an edge can in themselves have properties associated with them. In the SSSP problem setting, a vertex's distance can be viewed as a node property, being computed by the corresponding algorithm. Similarly, in the BC computation, the betweenness centrality values of each node can be viewed as a property. 
The \texttt{propNode} datatype facilitates declaring property for nodes of a graph with the provision of specifying its type. The \texttt{attachNodeProperty} function provided by the Graph type binds this property to the graph’s nodes and initializes the property values if provided. Line~\ref{computeSSSP-stat-3} in our SSSP code from Figure~\ref{sssp-dsl-sample} specifies the declaration of a node property \light{dist} of type \texttt{int}. The \texttt{attachNodeProperty} binds \light{dist} to the graph and optionally, initializes the distance attribute for each vertex (e.g., to infinity in Figure~\ref{sssp-dsl-sample}). Similarly, \texttt{propEdge} datatype is associated with edges, and has otherwise the same semantics as that of \texttt{propNode}. The \texttt{attachEdgeProperty} function binds the property to the graph’s edges. 

\name also provides collection types such as \texttt{List}, \texttt{SetN}, and \texttt{SetE}. \texttt{List} allows the presence of duplicates whereas \texttt{SetN} and \texttt{SetE} store unique nodes and edges respectively. 
Line~\ref{toy-stat-2} in Figure~\ref{toy-code} shows an example usage. 
The separation of sets between nodes and edges enables choosing the relevant implementation in vertex-based vs. edge-based codes. 

\subsubsection{Parallelization and Iteration Schemes}

\texttt{forall} is an aggregate construct in \name which can process a set of elements in parallel. Its sequential counterpart is a simple \light{for} statement. Currently, \name supports vertex-based processing\footnote{We are adding support for edge-based processing, which needs changes to the underlying data representation. Compressed Sparse Row (CSR) storage format is suited for vertex-based processing.}. 
The parallel \texttt{forall} supports various ranges it can iterate on (e.g., nodes in the whole graph or neighbors of a node, as shown in Figure~\ref{sssp-dsl-sample}, Lines~\ref{computeSSSP-stat-11} and \ref{computeSSSP-stat-13}). 


The function \texttt{g.nodes()} called on a graph \texttt{g} returns a sequence of nodes which can be iterated upon. To iterate over the neighbors of a node \texttt{u}, the functions \texttt{g.neighbors(u)}, \texttt{g.nodesTo(u)} and \texttt{g.nodesFrom(u)} return a similar sequence.
The \texttt{forall} body can be executed selectively for the nodes satisfying a certain boolean expression based on the node label or, node's property by including a \texttt{filter} construct. Line~\ref{toy-stat-3} in Figure~\ref{toy-code} shows its usage using the node property \texttt{modified}.


Considering that several graph algorithms can be well represented using a single outer parallel loop, and that the analysis of nested parallel loops gets complicated, currently, \name supports only outer level parallelism. Hence a nested \texttt{forall} in the DSL results in a parallel outer loop and a sequential inner loop in the target code. 
One may argue that an outer loop over vertices and an inner loop over neighbors can benefit from nested parallelism, and we agree. However, (i) such a processing can be well taken care of by an edge-based parallelism (to be supported), and (ii) since we target large graphs, even a vertex-based processing has enough parallelism to keep the resources busy.

For a graph processing DSL, traversals become the fundamental building blocks.
\name provides breadth-first traversal as a construct, borrowed from GreenMarl~\cite{GreenMarl}. 

\texttt{\keywordSP{iterateInBFS}(v \keywordSP{in} g.nodes() \keywordSP{from} root) \{...\}}

\noindent \texttt{iterateInBFS} performs a BFS traversal of the graph from the given root node. The underlying processing is level-by-level and iterates in parallel over the visited nodes in a specific level. On visiting a node in a level, it executes the body statements and forms the next set of visited nodes. The \texttt{filter()} construct can be utilized to explore the neighborhood of a visited node selectively.  Similarly, \texttt{iterateInReverse} performs a reverse BFS traversal in a level-synchronous manner and extracts parallelism at each level in the computation of the body statements. Note that \texttt{iterateInBFS} is a prerequisite to use \texttt{iterateInReverse}, since the former builds the BFS DAG to be traversed through in the latter. The functions \texttt{neighbors}, \texttt{nodesTo} and \texttt{nodesFrom} have a subtle change in their meaning when used inside \texttt{iterateIn...} constructs: they correspond to the neighbors in the BFS DAG rather than the original graph \texttt{G}. We have found this semantics change to satisfy our natural inclination to write a code. We illustrate it in the BC computation (Appendix~\ref{starplat:codes}).


\subsubsection{Reductions}
Reductions are one of the popular parallel programming primitives, and can be useful in achieving efficient computation. Specifying a reduction in the DSL also helps in conveying a necessity of synchronization. Unfortunately, it does not directly fit into the philosophy of \name design to support reduction as a language construct. Therefore, as a golden-mid, \name permits usage of certain relative C-operators (e.g., \texttt{+=}) to convey reduction. This "trick" allows us to retain the abstraction and still achieve efficiency of the generated code.
The reduction operators supported by \name are tabulated in Table~\ref{tab:reduction}.

\begin{table}[!htb]
\centering
\captionsetup{justification=centering}
\begin{center}
\begin{tabular}{ |c|c| } 
 \hline
 {\textbf{Operator}} & {\textbf{Reduction Type}}\\
 \hline
 {\texttt{+=}} & {Sum} \\
 \hline 
 {\texttt{*=}} & {Product} \\ 
 \hline 
 {\texttt{++}} & {Count} \\
 \hline 
  {\texttt{\&\&=}} & {All}\\
 \hline 
 {\texttt{||=}} & {Any} \\
 \hline 
 
\end{tabular}
\caption{Reduction operators in \name}
\label{tab:reduction}
\end{center}
\end{table}

We illustrate the usage of reduction in Figure~\ref{reduction-dsl}. The introduction of reduction (Line~\ref{reduction-accum} in the code makes sure the \texttt{accum} variable has a deterministic result at the end of the parallel region. Note that Line~\ref{reduction-count} involves a thread-local variable \texttt{count} and does not need reduction. On the other hand, if nested parallelism was supported, \texttt{count} would also need a reduction. The reduction operators in \name translate to library based implementations of reduction in the target backend.\footnote{Currently, the onus of writing the correct operator is onto the DSL user. We are adding program analysis to \name which would relieve the user of worrying about this.}

\begin{lstlisting}[language=NEAR, style=mystyle, label=reduction-dsl,  caption = Reduction example]
int accum = 0;
forall(v in g.nodes()){   
    int count = 0;
    forall(nbr in g.neighbors(v)){
           count = count + nbr.A;   // regular assignment /*@\label{reduction-count}*/
    }
   accum += count; //sum reduction in assignment form /*@\label{reduction-accum}*/
}
\end{lstlisting}

\subsubsection{fixedPoint and Min/Max Constructs}
Several solutions to graph algorithms are iterative, and converge based on conditions on node attributes. 
\name provides a fixedPoint construct to specify this succinctly. Its syntax involves a boolean variable and a boolean expression on node-properties forming the convergence condition, as shown below.

\texttt{\keywordSP{fixedPoint until} (var: convergence expression) \{...\}}

\noindent Line~\ref{computeSSSP-stat-9} of Figure~\ref{sssp-dsl-sample} in SSSP's specification  uses the \texttt{fixedPoint} construct to define the convergence condition. The loop iterates till at least one node's \texttt{modified} property is set to true. 

\name provides constructs \texttt{Min} and \texttt{Max} which perform multiple assignments based on a comparison criterion. This can be useful in update-based algorithms like SSSP, where an update on node properties is carried out on a  desired condition, while taking care of potential data races. 


In the SSSP computation, the \texttt{Min} construct is used to encode the relaxation criteria in Line~\ref{computeSSSP-stat-16}.
The neighboring node's distance is updated if the alternative distance via vertex \texttt{v} is smaller than \texttt{nbr}'s current distance. The update of the \texttt{dist} property based on this comparison specified using \texttt{Min} results in an update of the \texttt{modified} property to \texttt{True}. 

\name also has aggregate functions \texttt{minWt} and \texttt{maxWt} to find the minimum and maximum edge weights.

\subsection{Abstract Syntax Tree (AST)}

\begin{figure}
\centering
\includegraphics[scale=0.66]{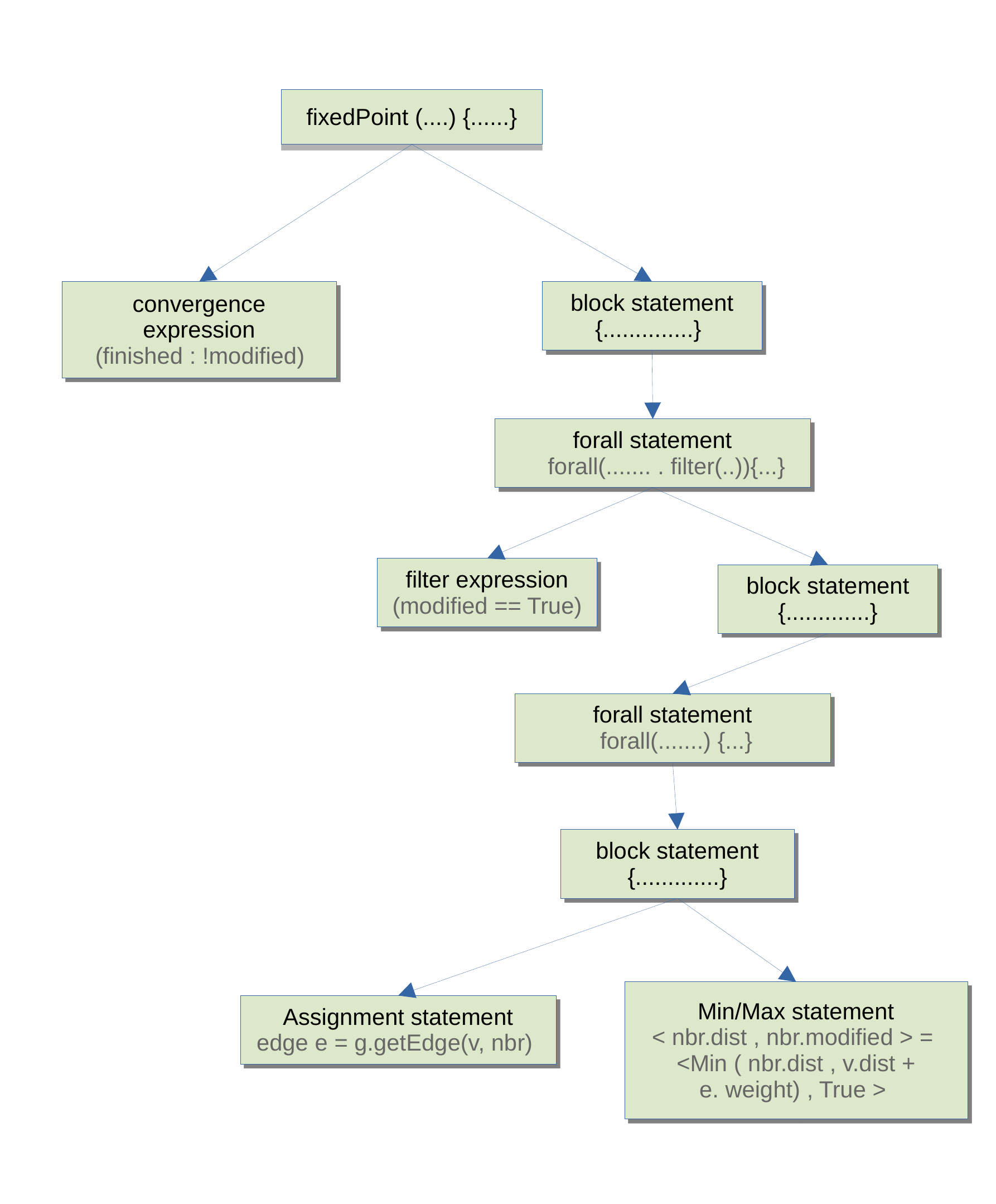}
\caption{AST for SSSP's \texttt{fixedPoint} construct in Figure~\ref{sssp-dsl-sample}} 
\label{fig:ast-fp}
\end{figure}

Each meaningful non-terminal that builds the language are at the highest level denoted as \texttt{ASTNode}. The
ASTNode is the parent class of all the nodes that are part of the abstract syntax tree. The
ASTNode forges to child class nodes that signify specialization like \texttt{statement} Node, \texttt{Expression} Node. The \texttt{statement} node is the parent class of all statement types like assignment, declaration, control flow, procedure call, etc. Each of these nodes can sometimes be  composed of other nodes, owing to the way the construct is defined in the parser. For instance, \texttt{forallStmt’s} construct is composed of a body, an iteration space which is defined by an iterator and range (functions) and an optional filter expression on the iterated items. The body in forallStmt is represented as a statement node, iterator as an \texttt{Identifier} node, ranges as \texttt{proc\_callExpr}, and filter expressions as \textit{Expression} node in the AST. The other node types are \texttt{Identifier}, \texttt{iterateReverseBFS}, \texttt{Type}, \texttt{reductionCall}, \texttt{fixedPointStmt}, \texttt{formalParam} etc. 

Most of the node's data are populated in the parsing stage itself. Data related to the type of the symbols are added during an additional pass through the already built AST. Given the target backend, the construct's AST are populated with additional information relevant for the code-generation in that specific backend. The analyzer phase before the code generation performs analysis for optimized and efficient code generation. Some of these optimizations are specific to the target backend. The separation of nodes
into different subclasses allows maintaining handlers for different node types in the code generator. The code-generator forms the backend part of the compiler. Unlike the frontend which was primarily target-independent, backend is tightly coupled with the target parallelization platform. 


%% file: 3GEN.tex
\section{\name Code Generator} \label{sec code gen}
The code generation phase supports the generation of efficient code for primarily three target backends,  shared memory, distributed, and many-core platforms.  The generated code for each backend also makes use of specific backend libraries to handle the parallel processing intuitively specified in the \name code. Shared memory implementation uses OpenMP as the threading library, distributed memory implementation uses MPI, and many-core implementation uses CUDA.

\subsection{Graph Representation and Storage} \label{codegen:graphrepresentation}
While there are several formats in which graphs can be stored, we preferred a format which:
\begin{itemize}
    \item can preferably work across all the backends
    \item works well with vertex-centric algorithms, common in graph processing
    \item can be easily split
\end{itemize}

The last preference is not only due to the MPI backend, but also due to our future plans to extend \name for multi-GPU and heterogeneous backends. Since offset-based formats fit our requirements, we had the options of compressed sparse row (CSR) and coordinate list (COO) formats, of which CSR fits the second preference. Therefore, we chose to use it. CSR is also popularly used in sparse graph processing frameworks. \name has a backend graph library which takes care of loading a graph from the input file, and storing it in the CSR format.

The OpenMP backend does not pose any issues with respect to storing the graphs. In case of the MPI backend, the graph is loaded by rank 0 process and distributed using a
simple block partitioning approach among the MPI processes, resulting in \textit{local} and \textit{remote} vertices for each process. The information of only the local vertices is available for direct access, while that of the remote vertices needs to be communicated with explicit send-receives. The MPI backend maintains a mapping between the local vertex id and its corresponding global vertex id, for correct processing and for sending and receiving data. This is because, for instance, a large global vertex id may still map to a local id of 0 in an MPI process. The CUDA backend needs to transfer the graph and the associated attributes fully from the host to the GPU (using multiple \texttt{cudaMemcpy} calls). Since the algorithms work with static graphs, only the modified attribute information needs to be brought back from the GPU to the host, at the end of the processing.

Seamless management of both the graph representation and the attribute storage across various backends relieves the domain-user of complex and erroneous graph handling.

\subsection{Overall Flow}
In the OpenMP backend, the outermost \texttt{forall} gets translated to have \texttt{\#pragma omp parallel for} as a header. Input graphs are parsed and stored as C++ classes in the generated code, while the nodes and the edges are represented as integers (for their ids). The node properties and edge properties are translated as arrays of types specified in the property declaration. We have found that generating OpenMP code is relatively easier compared to the other two backends due to (i) a single device,  (ii) shared memory paradigm wherein all the variables are present in the same memory address space for all the threads, and (iii) pragma based processing which does not need to change the sequential code.

The MPI backend generates code that follows a bulk-synchronous processing (BSP) style, which involves a computation step followed by a communication step in every iteration. During the computation step, various MPI processes execute the main-loop code, update the attributes of their local vertices, and mark those for sending in a send-buffer for the corresponding global vertices.  During the communication phase, processes receive these updates from other processes, convert those to their local vertices, and update the attributes. Similar to OpenMP, the node edge attributes / properties are translated as arrays.

Unlike OpenMP and MPI, the CUDA backend needs to convert the \texttt{forall} loop into a GPU kernel, which has a different scope. This makes certain variables unavailable, which need to be explicitly allocated (\texttt{cudaMalloc}) and transferred (\texttt{cudaMemcpy}). In addition, the \texttt{forall} loop body may contain local variables, which get translated to thread-local variables in CUDA. To distinguish between these two kinds of variables, the CUDA backend performs a rudimentary analysis of the AST.\footnote{A full-fledged program analysis of the algorithm at the AST level is an ongoing work, which involves identifying races and addressing those with proper synchronization or communication.} Apart from each outermost \texttt{forall}, the initialization to attribute values needs to be converted to a separate kernel (which is a loop in OpenMP and MPI). The kernel launch configuration is set to a fixed number of threads per block (1024 in the generated code, which performs the best on an average in our experiments) and the number of blocks proportional to the number of vertices.

Apart from this, the boilerplate code consists of calls to timing functions to time the initialization, the fixed-point processing, and the data transfer where applicable.

\subsection{Neighborhood Iteration}
Iterating over the neighborhood is typically nested inside iterating over the graph vertices (possibly, with a filter, as in Figure~\ref{sssp-dsl-sample}). In the case of the OpenMP backend, the generated code snippet is as below.

\begin{lstlisting}[language=C++, style=mystyle, label= nbritr-omp,  caption = OpenMP code  generated for neighborhood iteration]
#pragma omp parallel for
for (int v = 0; v < g.num_nodes(); v++) {
    if (!modified[v]) continue;
    for (int edge = g.indexofNodes[v]; edge < g.indexofNodes[v + 1]; edge++) {
        int nbr = g.edgeList[edge];
        ...
}   }
\end{lstlisting}

The MPI code is very similar, but goes over a predefined range based on the process rank. 

\begin{lstlisting}[language=C++, style=mystyle, label= nbritr-mpi,  caption = {MPI code  generated for neighborhood iteration}]
mpi::communicator world;
myrank = world.rank();
np = world.size();

part_size = ceil(g.num_nodes() / (float)np);
startv = myrank * part_size;
endv = startv + part_size - 1;
...
for (int v = startv; v <= endv; v++) {
    for (int edge0 = local_index[v-startv]; edge0 < local_index[v-startv+1]; edge0++) {
          int nbr = local_edgeList[edge0];
        ...
} }
\end{lstlisting}

The CUDA code needs to separate the kernel call and kernel processing. The kernel call sets up the launch configuration (number of thread-blocks and the size of each thread-block) and passes appropriate parameters (graph, attributes, algorithm specific variables such as the source node in SSSP, and other miscellaneous variables used internally). The kernel uses 1D thread id, and utilizes the CSR copied to the GPU to go over the neighbors. Note that unlike OpenMP and MPI, there is no loop over the vertices, which is handled by parallel threads with which the kernel is launched.

\begin{lstlisting}[language=C++, style=mystyle, label= nbritr-cuda,  caption = {CUDA code  generated for neighborhood iteration}]
__global__ void computeSSSP(...) {
    unsigned id = blockIdx.x * blockDim.x + threadIdx.x;
    ...
    for (int edge = gpu_OA[id]; edge < gpu_OA[id + 1]; edge++) {
        int nbr = g.gpu_edgeList[edge];
        ...
    }
    ...
}
...
computeSSSP<<<nblocks, blocksize>>>(...);
cudaDeviceSynchronize();

\end{lstlisting}

The loop bound functions change appropriately for directed vs. undirected graphs, and for in-neighbors (\texttt{g.nodesTo(v)}) which demand using reverse adjacencies (\texttt{g.revIndexofNodes[v]}).

\subsection{Reductions}
Recall the reduction supported in \name (e.g., Line~\ref{reduction-accum}). The reduction on a scalar translates to OpenMP's \texttt{reduction} clause as below. OpenMP reduction implementation creates and updates a private copy of the reduction variable per thread. At the exit of the parallel region,  the local changes are accumulated in the global variable. This significantly reduces the data race handling costs required for the correctness of operation on a shared variable. 

\texttt{\#pragma omp parallel for reduction(+:accum)}


\REM {
The MPI backend uses \texttt{MPI\_Reduce} to implement reduction. The function is called with a root process (id 0) and a reduction operation (MPI\_SUM in our example). The end result is stored in the root process. 
Similar to \texttt{Min/Max} on properties, every process iterates over its local vertices and computes the local reduction output. The global reduction is done using call to \texttt{MPI\_Reduce}. The generated MPI code is shown below.  \etodo{This reduction code won't be generated.. reduction has issues.}

\begin{lstlisting}[language=c++, style=cstyle, label= scalar-reduction-mpi, caption = MPI code generated for reduction (comments are not generated)]
int accum = 0;
for (int v = startv; v <= endv; v++) {
    for (int edge = g.indexofNodes[v]; edge < g.indexofNodes[v+1]; edge++) {
        int nbr = g.edgeList[edge];
        accum += prop[nbr];  // local accumulate
}   }
accum = all_reduce(world, accum, std::plus<int>()); // global reduce
\end{lstlisting}
}

The situation in CUDA gets complicated due to limitations on the number of resident thread-blocks, which can block the kernel. One option to counter this is to use a reduction from the host as a separate kernel (e.g., using \texttt{thrust::reduce}). But this demands terminating the kernel, coming back to the host, calling reduce, coming back to the host, and then calling another kernel to perform the rest of the processing in the \texttt{forall} loop. This not only adds complication to the code generation, but also makes the overall processing inefficient. 
Therefore, we rely on atomics to generate the functionally equivalent code (e.g., using \texttt{atomicAdd} in the example in Figure~\ref{reduction-dsl}). 

\texttt{atomicAdd(\&accum, prop[nbr]);}

\subsection{BFS Traversal}
The \texttt{iterateInBFS} construct gets expanded to a parallel BFS. 
This single construct expands to around 40 lines of OpenMP code, 60 lines of MPI code, and 40 lines of CUDA code. Therefore, we show and discuss only crucial parts of the snippets below.

The OpenMP backend maintains a vector of vectors for tracking vertices level-wise as shown in the code snippet in Figure~\ref{bfsAbs-gen}. The BFS has an outer \texttt{while} loop and two inner loops: one to go over the vertices in a level (this is parallelized at Line~\ref{line:bfsparallel}) and the other to add the explored vertices in a level to the appropriate level-vector (Line~\ref{line:bfsconcat}), which acts as a frontier. The parallel loop examines a vertex's neighbors and checks if their current distances are not set (value -1) and sets those to 1 + current vertex's distance. To avoid data races, this update needs to be done atomically using a CAS (Line~\ref{line:bfscas}).

\begin{lstlisting}[language=C++, style=cstyle, label= bfsAbs-gen,  caption = OpenMP code generated for the \texttt{iterateInBFS} construct]
levelNodes[phase].push_back(root); 
...
while (bfsCount > 0) {
    ...
    #pragma omp parallel for
    for(int l = 0; l < prev_count ; l++) { /*@\label{line:bfsparallel}*/
        int v = levelNodes[phase][l];
        ...
        dnbr = __sync_val_compare_and_swap(&bfsDist[nbr], -1, bfsDist[v] + 1);  /*@\label{line:bfscas}*/
        if (dnbr < 0) {
          int num_thread = omp_get_thread_num();
           levelNodes_later[num_thread].push_back(nbr);
        }
        // body of iterateInBFS
    }
    phase = phase + 1;

    for (int i = 0; i < omp_get_max_threads(); i++) {  /*@\label{line:bfsconcat}*/
        levelNodes[phase].concat(levelNodes_later[i]);
        ...
    }
}
\end{lstlisting}

The MPI backend runs a similar outer \texttt{while} loop and goes over the vertices level by level, but only for the set of local vertices. Once a vertex's neighbor is updated, it is pushed to the send buffer (Line~\ref{line:bfssendbuf}). At the end of each level, an all-to-all call exchanges the level values of local and remote vertices (Line~\ref{line:bfsalltoall}). Different levels are separated by an MPI barrier in a BSP style (Line~\ref{line:bfsbarrier}). 

\begin{lstlisting}[language=C++, style=cstyle, label= bfs-mpi, caption = MPI code generated for the \texttt{iterateInBFS} construct]
while(is_finished(startv, endv, active)) {
    ...
    while (active.size() > 0) {
        int v = active.back();
        active.pop_back();
        ...
        for (int j = local_index[v - startv]; j < local_index[v - startv + 1]; j++) {
            int w = local_edgeList[j];
            //If local 
            if (d[w-startv] < 0) {
                active_next.push_back(w);
                d[w-startv] = d[v-startv] + 1;
            } else 
                // send d[v], w, and v /*@\label{line:bfssendbuf}*/
        ...
    }
    all_to_all(world, sendbuf, receivebuf); /*@\label{line:bfsalltoall}*/
    for (int t = 0; t < np; t++) {
        // process receives
        if (d[w-startv] == d_v + 1)
            //Body of iterateInBFS
    }
    MPI_Barrier(MPI_COMM_WORLD); /*@\label{line:bfsbarrier}*/
    phase = phase + 1 ;
}
\end{lstlisting}

The CUDA backend poses a challenge for the \texttt{iterateInBFS} construct because unlike the other two backends, the code needs to be generated to be running on both the host and the device. The outer \texttt{do-while} loop runs on the host (Line~\ref{line:bfsdowhile}), which internally calls the level-wise BFS kernel on the GPU (Line~\ref{line:bfskernel}). Since the loop is on the host, for ensuring progress with the loop's condition, the flag (\texttt{finished}) needs to be repeatedly copied across devices (Lines~\ref{line:bfsh2d} and \ref{line:bfsd2h}), passed as a parameter to the kernel, and updated in the kernel whenever there is change of a vertex's level.   

\begin{lstlisting}[language=NEAR, style=cstyle, label=bfs-cuda, caption = CUDA equivalent code for the \texttt{iterateBFS} construct]
do {          /*@\label{line:bfsdowhile}*/
    H2D(...); /*@\label{line:bfsh2d}*/

    BFS<<<...>>>(...)
    cudaDeviceSynchronize();
    ++hops_from_source;

    D2H(...); /*@\label{line:bfsd2h}*/
    ...

} while (!finished);

__global__ void BFS(...) { /*@\label{line:bfskernel}*/
    ... 
    if (d_level[u] == *d_hops_from_source) {
        ... 
        for (unsigned i = d_offset[u]; i < end; ++i) {
            unsigned v = ... if (d_level[v] == -1) {
                d_level[v] = *d_hops_from_source + 1;
                *d_finished = false;
            }
            // Body of iterateInBFS
}   }   }
\end{lstlisting}

\subsection{\texttt{Min}/\texttt{Max} Constructs}
Recall the \texttt{Min} construct from SSSP (Line~\ref{computeSSSP-stat-16}). In OpenMP, the conditional update on the node property is achieved through  atomic implementations of these operations built upon atomic CAS (Compare and Swap), 
made available as a \name library. The generated OpenMP code for the \texttt{Min} construct in SSSP is as follows.

\begin{lstlisting}[language=C++, style=cstyle, label= minred-gen,  caption = {OpenMP code generated for the \texttt{Min} construct}]
int dist_new = dist[v] + weight[e];
bool modified_new = true;

if (dist[nbr] > dist_new) {
    int oldValue = dist[nbr];
    atomicMin(&dist[nbr], dist_new);

    if(oldValue>dist[nbr] {
        modified_nxt[nbr] = modified_new;
}   }
\end{lstlisting}

In the MPI backend, the neighbor could be local or remote. If it is local, each process can update its property without synchronization. If it is remote, the property value to be updated is stored in the send buffer, which is eventually sent to the owning process (during the communication phase). 

\begin{lstlisting}[language=C++, style=cstyle, label= reduction-mpi, caption = {MPI code generated for the \texttt{Min} construct (comments are not generated)}]
if (nbr >= startv && nbr <= endv) {	// local
    int e = edge0 + startv;
    int dist_new = dist[v-startv] + weight[e-startv];
    bool modified_new = true;
	
    if (dist[nbr-startv] > dist_new) {
        dist[nbr-startv] = dist_new;
        modified[nbr-startv] = modified_new;
    }
} else {  // remote
    sendbuf[owner(nbr)][nbr] = dist[v-startv] + weight[e-startv];
}
...
all_to_all(world, sendbuf, receivebuf);

\end{lstlisting}

The CUDA backend handles Min/Max constructs exactly similar to the OpenMP backend, except (i) the variables used are the GPU copies, and (ii) the atomic instructions are readily supported in CUDA (\texttt{atomicMin} and \texttt{atomicMax}).

\begin{lstlisting}[language=C++, style=cstyle, label=reduction-cuda, caption = CUDA code generated for the \texttt{Min} construct]
int nbr = gpu_edgeList[edge];
int e = edge;
int dist_new = gpu_dist[v] + gpu_weight[e];

if (gpu_dist[nbr] > dist_new) {
    atomicMin(&gpu_dist[nbr], dist_new);
    gpu_modified_next[nbr] = true;
    gpu_finished[0] = false;
}
\end{lstlisting}

\subsection{\texttt{fixedPoint} Construct}
The \texttt{fixedPoint} construct translates to a while loop conditioned on the fixed-point variable provided in the construct. 
Typically, the convergence is based on a node property, but it can be an arbitrary computation. 
This code is generated on the host, so it is similar in template for all the three backends. However, as before, the MPI code needs to gather this information for setting the flag \texttt{finished} from all the processes based on the vertex ids they operate on. Similarly, the CUDA code updates a copy of the \texttt{finished} flag on the GPU, which is \texttt{cudaMemcpy}'ed to the host.

\begin{lstlisting}[language=C++, style=cstyle, label= fixedPoint-gen,  caption = Code generated for the \texttt{fixedPoint} construct across all the backends]
while (!finished) {
    finished = true;

    if (...) { // Min-construct expansion
        ...        
        finished = false; // fixedPoint identifier updation
}   }   
\end{lstlisting}

%% file: 4OPT.tex
\section{Optimizations} \label{sec optimizations}
There are certain optimizations which a hand-crafted code can embed, which \name's code generator cannot do. For instance, if variants of an algorithm perform better on different backends, we cannot auto-generate the variants from the same algorithmic specification (e.g., push versus pull based processing~\cite{pushpull}), although the DSL is capable of specifying the variants separately. At the implementation level, persistent kernels in CUDA~\cite{persistentkernels} can improve execution time of algorithms on large diameter graphs. However, \name generates the code for the \texttt{fixedPoint} construct on the host -- across backends. Despite this, our observation has been that many optimizations can be readily embedded into the code generated by \name:
\begin{itemize}
    \item The algorithmic optimizations can be embedded into the DSL specification itself, which forms a common input to all the backends. As an illustration, Figure~\ref{sssp-pull-dsl} in the Appendix presents a pull-based SSSP.
    \item A backend-specific optimization can be embedded into that specific backend, since such an optimization will not be explicitly mentioned in the DSL specification and will be abstracted away.
\end{itemize}

We discuss below the backend-specific optimizations.

\subsection{OpenMP} \label{subsec opt omp}
\mypara{Avoiding false sharing in \texttt{iterateInBFS}} In the case of BFS abstractions, nodes discovered by multiple threads at a particular level are pushed to a data structure required for graph traversal. The false sharing due to shared usage of the same array in this scenario has been alleviated by assigning local update space to each thread. 

\mypara{Efficient fixed-point computation} The fixedPoint construct converges on a specific condition on a single boolean node property. The change of convergence is tracked through a boolean fixed-point variable ideally needs to be updated after analyzing the property values for all nodes. The update procedure has been optimized by updating the fixed-point variable along with the update to the property value for any node. Since, updates to a boolean variable by multiple threads are atomic by hardware, this does not lead to a performance loss.

\mypara{Using built-in atomics} Atomics has proved to be lightweight. Hence, atomics implemented using built-in functions provided by the GCC compiler is being used in generated code for achieving exclusive access to a single statement by a thread instead of locks.

\subsection{MPI} \label{subsec opt mpi}
\mypara{Communication aggregation} The communication aggregation optimization is used by the MPI code while updating the distance of remote neighbor vertices. Instead of sending multiple messages to a remote vertex by a processing node, a single message with local minimum value is sent to the remote vertex. 

\mypara{Using Boost} The packing/unpacking mechanism provided by Boost MPI library gives better performance in sending and receiving STL containers such as vectors, maps etc., and user-defined data types than MPI library calls.

\mypara{Quick index-based partitioning} \name currently uses an index based partitioning to distribute the graph among various MPI processes. Compared to using a partitioner such as Metis, ours has the down-side of increased communication due to several inter-partition edges. However, such a partitioning is quick, and also allows us to partition the graph arbitrarily based on the number of MPI processes. Presently, the implementation assumes that all the processes have equal number of vertices. Hence, we pad temporary vertices for the last process.\footnote{We definitely plan to exploit Metis and its parallel variants for MPI, multi-GPU, and heterogeneous backends.}

\subsection{CUDA} \label{subsec opt cuda}
\mypara{Optimized host-device transfer} 
We perform a rudimentary program analysis of the AST to identify variables that need to be transferred across devices. For instance, since graph is static, it need not be copied back from GPU to CPU at the end of the kernel. In contrast, the modified properties need to be transferred back. Similarly, the \texttt{finished} flag is set on CPU, conditionally set on GPU, and read on the CPU again in the fixed-point processing. Therefore, the variable needs to be transferred to-and-fro. The \texttt{forall}-local variables are generated as device-only variables.

\mypara{Memory optimization in OR-reduction} The way we write the \texttt{fixedPoint} construct, a \texttt{modified} property is used in computing the fixed-point. At a high-level, another iteration is necessary if any of the vertices' \texttt{modified} flag is set. This is essentially a logical-OR operation. \name takes advantage of this to generate a single flag variable which is set by threads in parallel (with dependence on hardware atomicity for primitive types). Managing this flag is cheaper than transferring arrays of the \texttt{modified} flags across devices, both in terms of time and memory. 

%% file: 5EXP.tex

\section{Experimental Evaluation} \label{sec exp evaluation}
To quantitatively assess \name, we generate four popular algorithm implementations: Betweenness Centrality (BC), PageRank (PR), Single Source Shortest Path (SSSP), and Triangle Counting (TC). These are also coded and optimized using the other frameworks we compare against. The \name code for SSSP is presented in Figure~\ref{sssp-dsl-sample}, while the other three are presented in the Appendix. Each codes fits in about 30 lines, and with that effort, a domain-expert or a student can generate efficient implementations of the four algorithms for a multi-core setup, a distributed system, and a GPU.\footnote{We believe this is remarkable, and would be appreciated by the community.} 

Since the DSL codes are rather short, the compilation is immediate.
Therefore, we focus on analyzing the efficiency of the generated codes for the various backends. We compare the performance of the backend-specific code in \name against the existing state-of-the-art graph analytic solutions (qualitative comparison is discussed in Section~\ref{sec relatedwork}). \name's OpenMP backend is compared against Galois~\cite{Galois}, Green-Marl~\cite{GreenMarl}, and Ligra~\cite{Ligra}; its MPI backend is compared against Gluon~\cite{Gluon-DGalois}; and its CUDA backend is compared against GunRock~\cite{Gunrock} and LonestarGPU~\cite{lonestargpu}. Galois is a C++-based framework for graph analytics, while Ligra is C-based. In addition, Galois also supports graph mutation, which is not the focus of other frameworks. Similar to \name, Green-Marl is a graph DSL designed for the multi-core backend. Gluon is a distributed version of Galois built as a C++ framework. LonestarGPU is a manually optimized collection of CUDA codes, while Gunrock is a CUDA-based library providing data-centric API for graph analytics, such as \textit{advance}, \textit{compute}, and \textit{filter}. So, except for Green-Marl, we compare \name-generated codes against manually optimized ones. 

We use ten large graphs in our experiments, which are a mix of different types. Six of these are social networks exhibiting the small-world property, two are road networks having large diameters and small vertex degrees, while two are synthetically generated. One synthetic graph has a uniform random distribution (generated using Green-Marl's graph generator), while the other one has a skewed degree distribution following the recursive-matrix format (generated using SNAP’s RMAT generator with parameters a = 0.57, b = 0.19, c = 0.19, d = 0.05). They are listed in Table~\ref{graph-inputs}, sorted on the number of edges in each category. 
For unweighted graphs, we assign edge-weights selected uniformly at random in the range [1,100] (for SSSP).

\begin{table}[H]
\centering
\begin{tabular}{ r|c|r|r|r|r|r }
 \hline
 \textbf{Graph} & \textbf{Acronym} & \textbf{Num. Vertices} &  \textbf{Num. Edges}  
 & \textbf{Avg. Degree} & \textbf{Max. Degree} \\
    &  & \multicolumn{1}{c|}{(million)} & \multicolumn{1}{c|}{(million)} & & & \\
 \hline
 twitter-2010 & TW   & 21.2  & 265.0 & 
 12 & 302,779\\ 
 soc-sinaweibo & SW &   58.6  & 261.0 & 
 4 & 4,000 \\
 orkut & OK & 3.0 & 234.3 
 & 76.2813  &33,313\\
 wikipedia-ru & WK & 3.3 & 93.3 
 & 55.4067 & 283,929\\
 livejournal &LJ &  4.8 & 69.0 
 &28.257 & 22,887\\
 soc-pokec & PK & 1.6 & 30.6 
 & 37.5092 & 20,518\\
 \hline
 usaroad & US & 24.0  & 28.9 
 & 2 & 9  \\
 germany-osm & GR & 11.5  & 12.4 
 & 2  & 13\\ \hline
 rmat876 & RM & 16.7 & 87.6  
 & 5 & 128,332\\
 uniform-random & UR & 10.0 & 80.0 
 & 8 & 27 \\
 \hline
\end{tabular}
\caption{Input graphs} 

\label{graph-inputs}
\end{table}

All our experiments, including the baselines we compare against, were run on IIT Madras AQUA cluster. The configuration of each compute node as follows: Intel Xeon Gold 6248 CPU with 40 hardware threads spread over two sockets, 2.50~GHz clock, and 192 GB memory running RHEL 7.6 OS. 
All the codes in C++ are compiled with GCC 9.2, with optimization flag -O3. Various backends have the following versions: OpenMP version 4.5, OpenMPI version 3.1.6, CUDA version 10.1.243 
and run on Nvidia Tesla V100-PCIE GPU with 5120 CUDA cores spread uniformly across 80 SMs clocked at 1.38~GHz with 32~GB global memory and 48 KB shared memory per thread-block. 

\subsection{OpenMP} \label{subsec exp eval omp}
Table~\ref{openmp-table} presents the running times of the codes in various frameworks for the four algorithms. The execution time refers to only the algorithmic processing and excludes the graph reading time, and is an average over three runs (for each backend).
The Galois framework failed to load the largest graph TW (exited with a segfault). The framework also failed for BC computation on 150 sources for US graph and exited by giving a PTS out of memory error.
\arrayrulecolor{gray}
\begin{table}
\scalebox{0.9}{
\setlength{\tabcolsep}{2pt}
\begin{tabular}{|rr|r||r|r|r|r|r|r|r|r|r|r||r|}
\hline
\multicolumn{2}{|c|}{Algo.} & Framework & TW  & SW  & OK  & WK & LJ & PK & US & GR & RM & UR & Total\\ \hline
\multicolumn{1}{|r|}{\multirow{3}{*}{BC}} & 20  & \begin{tabular}[c]{@{}l@{}}Galois\\ Ligra\\ GreenMarl\\ \name\end{tabular} & \begin{tabular}[r]{@{}l@{}}-\\ 11.200\\ \textbf{4.289}\\ 6.530\end{tabular}               & \begin{tabular}[r]{@{}l@{}}\textbf{1.840}\\ 13.130\\ 3.492\\ 11.060\end{tabular}        & \begin{tabular}[r]{@{}l@{}}8.272\\ 5.936\\ \textbf{5.504}\\ 7.650\end{tabular}     & \begin{tabular}[r]{@{}l@{}}3.377\\ 3.470\\ \textbf{2.444}\\ 3.300\end{tabular}          & \begin{tabular}[r]{@{}l@{}}3.209\\ 4.060\\ \textbf{2.218}\\ 4.422\end{tabular}   & \begin{tabular}[r]{@{}l@{}}1.477\\ \textbf{1.220}\\ 1.173\\ 1.777\end{tabular}    & \begin{tabular}[r]{@{}l@{}}38.488\\ \textbf{18.160}\\ 18.823\\ 22.250\end{tabular} & \begin{tabular}[r]{@{}l@{}}15.368\\ 10.430\\ \textbf{9.602}\\ 11.740\end{tabular}    & \begin{tabular}[r]{@{}l@{}}\textbf{2.861}\\ 4.250\\ 3.187\\ 5.580\end{tabular} & \begin{tabular}[r]{@{}l@{}}29.528\\ 4.736\\ \textbf{3.573}\\ 12.830\end{tabular}
  & \begin{tabular}[r]{@{}l@{}}104.420\\ 76.592\\ 54.305\\ 87.139 \end{tabular} \\ 
\cline{2-14} 
\multicolumn{1}{|l|}{}  & 80  & \begin{tabular}[c]{@{}l@{}}Galois\\ Ligra\\ GreenMarl\\ \name\end{tabular} & \begin{tabular}[c]{@{}l@{}}-\\ 48.700\\    \textbf{18.189}\\ 33.240\end{tabular}           & \begin{tabular}[c]{@{}l@{}}\textbf{7.269}\\ 48.760\\ 12.172\\ 44.270\end{tabular}      & \begin{tabular}[c]{@{}l@{}}50.171\\ 24.100\\ \textbf{21.094}\\ 32.150\end{tabular}   & \begin{tabular}[c]{@{}l@{}}13.548\\ 14.230\\ \textbf{9.353}\\ 13.050\end{tabular}      & \begin{tabular}[c]{@{}l@{}}12.139\\ 16.160\\ \textbf{8.578}\\ 16.550\end{tabular} & \begin{tabular}[c]{@{}l@{}}5.596\\ 4.550\\ \textbf{4.107}\\ 6.810\end{tabular}     & \begin{tabular}[c]{@{}l@{}}258.447\\ \textbf{73.230}\\ 75.450\\ 86.400\end{tabular}  & \begin{tabular}[c]{@{}l@{}}75.969\\ 40.500\\ \textbf{39.704}\\ 47.290\end{tabular}    & \begin{tabular}[c]{@{}l@{}}\textbf{7.435}\\ 10.900\\ 7.863\\ 18.030\end{tabular}        & \begin{tabular}[c]{@{}l@{}}85.534\\ 19.130\\ \textbf{14.184}\\ 50.260\end{tabular}
 & \begin{tabular}[r]{@{}l@{}}516.108\\ 300.260\\ 210.690\\ 348.050 \end{tabular} \\ 
\cline{2-14} 
\multicolumn{1}{|l|}{}                    & 150 & \begin{tabular}[c]{@{}l@{}}Galois\\ Ligra\\ GreenMarl\\ \name\end{tabular} & \begin{tabular}[c]{@{}l@{}}-\\ 83.400\\ \textbf{32.990}\\ 61.625\end{tabular}           & \begin{tabular}[c]{@{}l@{}}\textbf{14.598}\\ 90.360\\ 23.110\\ 84.430\end{tabular}      & \begin{tabular}[c]{@{}l@{}}78.632\\ 44.630\\ \textbf{38.612}\\ 57.664\end{tabular} & \begin{tabular}[c]{@{}l@{}}25.403\\ 26.260\\ \textbf{17.709}\\ 24.520\end{tabular}     & \begin{tabular}[c]{@{}l@{}}23.816\\ 31.500\\ \textbf{17.231}\\ 32.290\end{tabular} & \begin{tabular}[c]{@{}l@{}}11.132\\ 8.996\\ \textbf{8.184}\\ 13.553\end{tabular} & \begin{tabular}[c]{@{}l@{}}-\\ \textbf{124.66}\\ 139.090\\ 160.865\end{tabular}   & \begin{tabular}[c]{@{}l@{}}138.052\\ 76.530\\ \textbf{74.506}\\ 80.111\end{tabular} & \begin{tabular}[c]{@{}l@{}}12.365\\ 17.700\\ \textbf{11.069}\\ 32.136\end{tabular}     & \begin{tabular}[c]{@{}l@{}}116.700\\ 35.800\\ \textbf{26.186}\\ 94.540\end{tabular}  & \begin{tabular}[c]{@{}l@{}}420.698\\ 539.836\\ 388.670\\ 641.734\end{tabular} \\ \hline
\multicolumn{2}{|l|}{PR}                             & \begin{tabular}[c]{@{}l@{}}Galois\\ {\color{red}{Ligra}}\\ GreenMarl\\ \name\end{tabular} & \begin{tabular}[c]{@{}l@{}}-\\ 25.600\\ \textbf{0.585}\\ 1.752\end{tabular}          & \begin{tabular}[c]{@{}l@{}}\textbf{0.510}\\ 162.660\\ 7.211\\ 9.002\end{tabular}      & \begin{tabular}[c]{@{}l@{}}\textbf{0.647}\\ 5.050\\ 1.437\\ 1.213\end{tabular}     & \begin{tabular}[c]{@{}l@{}}\textbf{0.371}\\ 3.930\\ 0.512\\ 0.473\end{tabular}         & \begin{tabular}[c]{@{}l@{}}\textbf{0.474}\\ 3.623\\ 0.585\\ 0.509\end{tabular}   & \begin{tabular}[c]{@{}l@{}}\textbf{0.156}\\ 0.836\\ 0.263\\ 0.236\end{tabular}   & \begin{tabular}[c]{@{}l@{}}\textbf{0.607}\\ 2.050\\ 1.235\\ 1.600\end{tabular}      & \begin{tabular}[c]{@{}l@{}} \textbf{0.224}\\ 0.822\\ 0.525\\ 0.667\end{tabular}     & \begin{tabular}[c]{@{}l@{}}\textbf{0.324}\\ 5.880\\ 0.821\\ 0.883\end{tabular} & \begin{tabular}[c]{@{}l@{}}\textbf{0.443}\\ 0.942\\ 0.688\\ 0.619\end{tabular}  & \begin{tabular}[c]{@{}l@{}} 3.756 \\ 211.393 \\ 13.862 \\ 16.954 \end{tabular} 
\\ \hline
\multicolumn{2}{|l|}{SSSP}                           & \begin{tabular}[c]{@{}l@{}}Galois\\ Ligra\\ GreenMarl\\ \name\end{tabular} & \begin{tabular}[c]{@{}l@{}}\textbf{0.522}\\ 10.7\\ 2.182\\ 5.831\end{tabular}          & \begin{tabular}[c]{@{}l@{}} \textbf{0.132}\\ 0.148\\ 0.891\\ 1.437\end{tabular}      & \begin{tabular}[c]{@{}l@{}}\textbf{0.404}\\ 5.136\\ 1.048\\ 1.850\end{tabular}    & \begin{tabular}[c]{@{}l@{}}\textbf{0.203}\\ 1.846\\ 1.16\\ 1.759\end{tabular}       & \begin{tabular}[c]{@{}l@{}}\textbf{0.205}\\ 3.89\\ 0.761\\ 2.412\end{tabular}  & \begin{tabular}[c]{@{}l@{}}\textbf{0.099}\\ 1.683\\ 0.292\\ 0.846\end{tabular}   & \begin{tabular}[c]{@{}l@{}}\textbf{19.387}\\ 283.000\\ 193.548\\ {\color{red}{294.303}}\end{tabular}   & \begin{tabular}[c]{@{}l@{}}\textbf{6.798}\\ 9.043\\ 48.349\\ {\color{red}{53.395}}\end{tabular}    & \begin{tabular}[c]{@{}l@{}}\textbf{0.133}\\ 2.74\\0.464\\ 1.369\end{tabular}       & \begin{tabular}[c]{@{}l@{}}\textbf{0.480}\\ 10.400\\ 1.361\\ 5.237\end{tabular}  & \begin{tabular}[c]{@{}l@{}} 28.363\\ 328.586\\ 250.056 \\ 368.439 \end{tabular} 
\\ \hline
\multicolumn{2}{|l|}{TC}                             & \begin{tabular}[c]{@{}l@{}}Galois\\ Ligra\\ {\color{red}{GreenMarl}}\\ \name\end{tabular} & \begin{tabular}[c]{@{}l@{}}-\\ 2103.333\\  11611.029\\ 1414.323\end{tabular} & \begin{tabular}[c]{@{}l@{}}\textbf{56.432}\\ 188.660\\ 4257.498\\ 59.925\end{tabular} & \begin{tabular}[c]{@{}l@{}}33.110\\ \textbf{22.800}\\ 137.559\\ 23.420\end{tabular}   & \begin{tabular}[c]{@{}l@{}}\textbf{46.168}\\ 97.360\\ 4564.568\\ 111.430\end{tabular} & \begin{tabular}[c]{@{}l@{}}9.811\\ 10.460\\ 29.426\\ \textbf{7.544}\end{tabular} & \begin{tabular}[c]{@{}l@{}}3.008\\ 1.926\\ 12.705\\ \textbf{1.559}\end{tabular}  & \begin{tabular}[c]{@{}l@{}}0.061\\ 0.147\\ 0.065\\ \textbf{0.059}\end{tabular}   & \begin{tabular}[c]{@{}l@{}}\textbf{0.020}\\ 0.0698\\ 0.021\\ 0.024\end{tabular}     & \begin{tabular}[c]{@{}l@{}}184.260\\ \textbf{130.330}\\ 5647.156\\ 158.760\end{tabular} & \begin{tabular}[c]{@{}l@{}}2.350\\ 1.706\\ 1.435\\ \textbf{1.176}\end{tabular}  & \begin{tabular}[c]{@{}l@{}}335.220\\ 2556.792\\ 14650.430\\ 1778.220\end{tabular}     
\\ \hline
\end{tabular}
}
\caption{\name's OpenMP code performance comparison against Galois, Ligra and Green-Marl (20 Threads). All times are in seconds. BC is run with the number of sources as \{20, 80, 150\}.}
\label{openmp-table}
\end{table}

\begin{table}
\begin{tabular}{|l|l|l|l|l|l|l|l|l|l|l|}
\hline
\multirow{2}{*}{SSSP} & TW   & SW    & OK    & WK    & LJ    & PK    & US   & GR    & RM    & UR    \\ \cline{2-11} 
                      & 4.127 & 0.127 & 1.503 & 0.633 & 2.315 & 0.822 & 72.654 & 9.641 & 1.319 & 4.477 \\ \hline
\end{tabular}
\caption{\name's SSSP OpenMP code running times (seconds) with \texttt{static} scheduling}
\label{openmp-dyn-table}
\end{table}

\mypara{\bcc} BC resembles all-pairs shortest paths, which can be implemented by running SSSP from each vertex as a source. Complete execution of BC on large graphs takes several hours, sometimes days. Depending upon how an implementation stores BC results for each source, the memory requirement can also increase proportional to $\sim V\sim$ per source. Therefore, literature presents results of running BC from one or only a few vertices. We tabulate results for BC for number of source vertices as \{20, 80, 150\}. The source list for each graph is generated using a random number generator and fixed across frameworks for consistent comparison. \name's BC code outperformed Galois code for the road network graphs and some of the social network graphs. Galois framework involves a scheduling policy that ensures better load balance among threads. But for jobs with even workloads, the scheduling adds to the runtime overhead. Ligra has an optimized forward BFS pass that bypasses distance computation for nodes, and computes only the number of shortest paths for each node, thus saving on the synchronization required for concurrent distance computation. This optimization results in Ligra outperforming \name and Galois. Interestingly, the Green-Marl-generated code outperforms the other three frameworks on most inputs. 
Green-Marl's forward BFS traversal is optimized and chooses the nature of traversal at each level based on specific criteria.

\mypara{\prr} \name-generated PR code is competitive to that of Green-Marl across inputs. Ligra, interestingly, has considerably slower implementation and takes significantly longer. This is primarily due to the loop separation between the computation of the PR values and the difference of successive PR values for each vertex.  Overall, Galois (about 2$\times$ faster than \name-generated code) outperforms all the other benchmarks. This is due to the in-place update of the PR values for vertices, which leads to faster convergence. \name, Green-Marl, and Ligra follow a similar processing of updating the PR values using double buffering. 

\mypara{\ssspp} We find that  Galois SSSP is faster than \name's code and other compared benchmarks. This is due to application-specific prioritized scheduling in the Galois framework~\cite{galois-scheduling}. For instance, processing tasks in the ascending distance order reduces the total amount of extra work done. GreenMarl and StarPlat have nearly similar implementations for SSSP calculation. Both follow a dense push configuration for vertex processing which requires iterating over all the vertices to check if they are active. This is typically costly for road networks which have a smaller frontier active in each iteration. In addition, this additional cost accumulates over several iterations due to the large diameter of road networks. However, GreenMarl performs better than StarPlat for nearly all graphs. GreenMarl uses spin-lock implementation with the back-off strategy to save on unnecessary CPU cycles. StarPlat also uses a lock-free atomic-based implementation but the unnecessary updates are more profound leading to performance degradation due to false sharing. Ligra's performance is not very competitive with other benchmarks. Ligra switches between sparse and dense edge processing based on the frontier size. But this direction optimization does not lead to considerable performance improvement for the given graph suite except for the road networks. 

 By default, \name generates OpenMP code with \texttt{dynamic} scheduling. This largely works well across various algorithms and graph types. However, SSSP code seems to overall perform better with static scheduling as shown in Table~\ref{openmp-dyn-table}. The difference is pronounced for large diameter graphs US and GR wherein the execution times reduce from over a minute to a few seconds. 

\mypara{\tcc} Galois, \name, and Green-Marl follow a node-iterator pattern in TC. On the other hand, Ligra follows an edge-iterator based version, which is supposed to work better for skewed degree graphs, since the edge-based version has better load balance. We observe that for different graphs, different frameworks outperform. Interestingly, performance of the Green-Marl-generated code is significantly poorer. 


\begin{figure}
    \begin{subfigure}[b]{0.49\textwidth}
    \centering
    \includegraphics[scale=0.5]{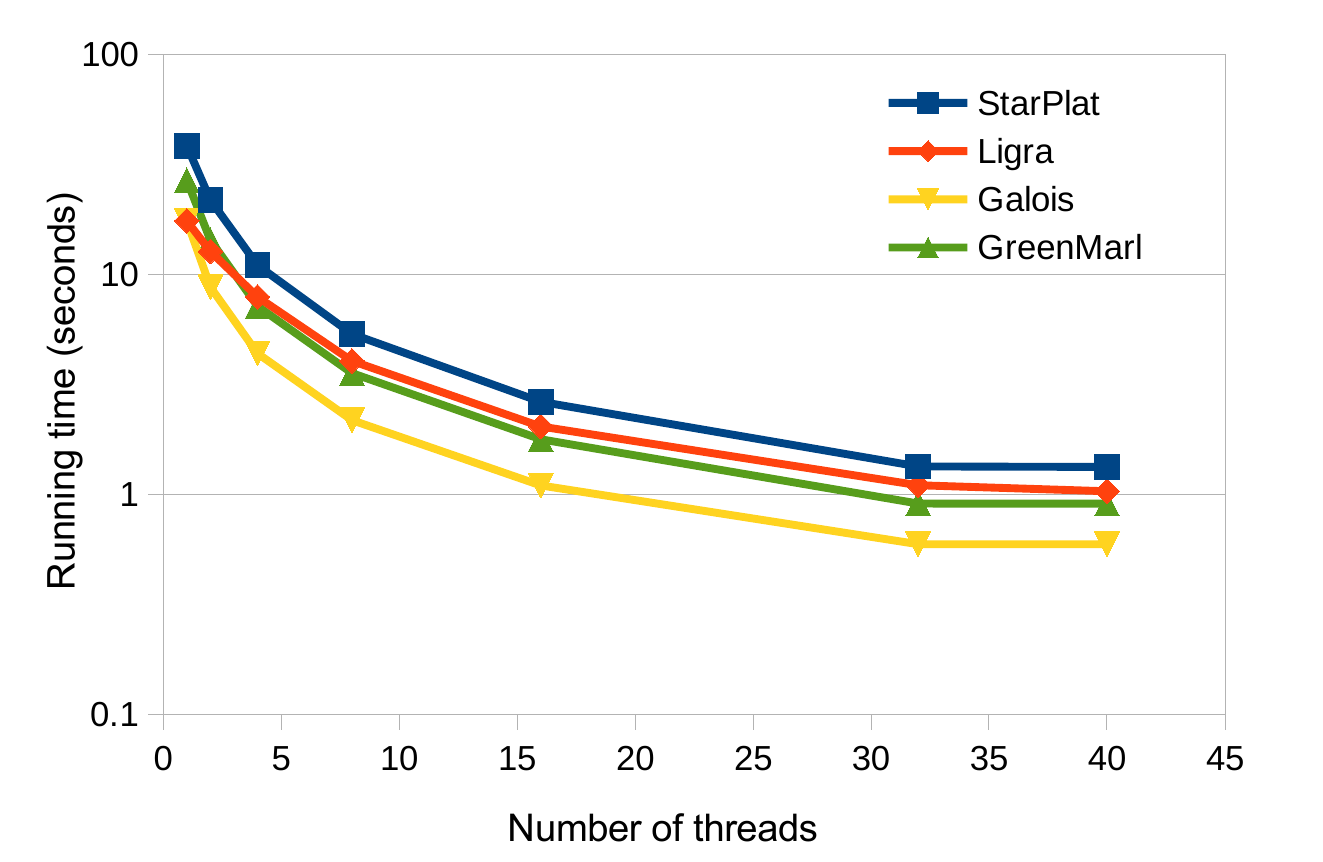}
     \caption{UR}
    \end{subfigure}
    \begin{subfigure}[b]{0.49\textwidth}
    \centering
    \includegraphics[scale=0.5]{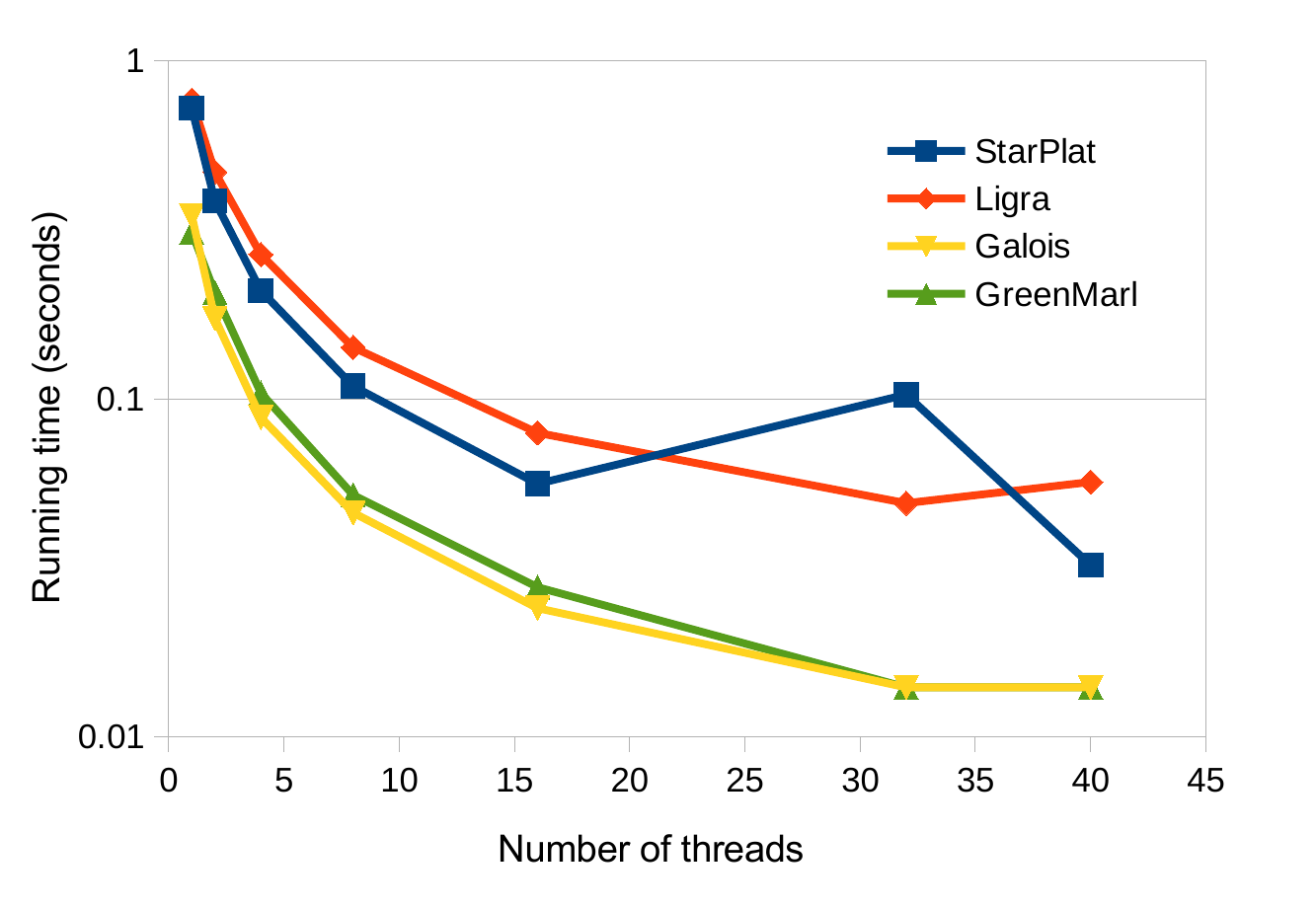}
    \caption{GR}
    \end{subfigure}
    \begin{subfigure}[b]{0.49\textwidth}
    \centering
    \includegraphics[scale=0.5]{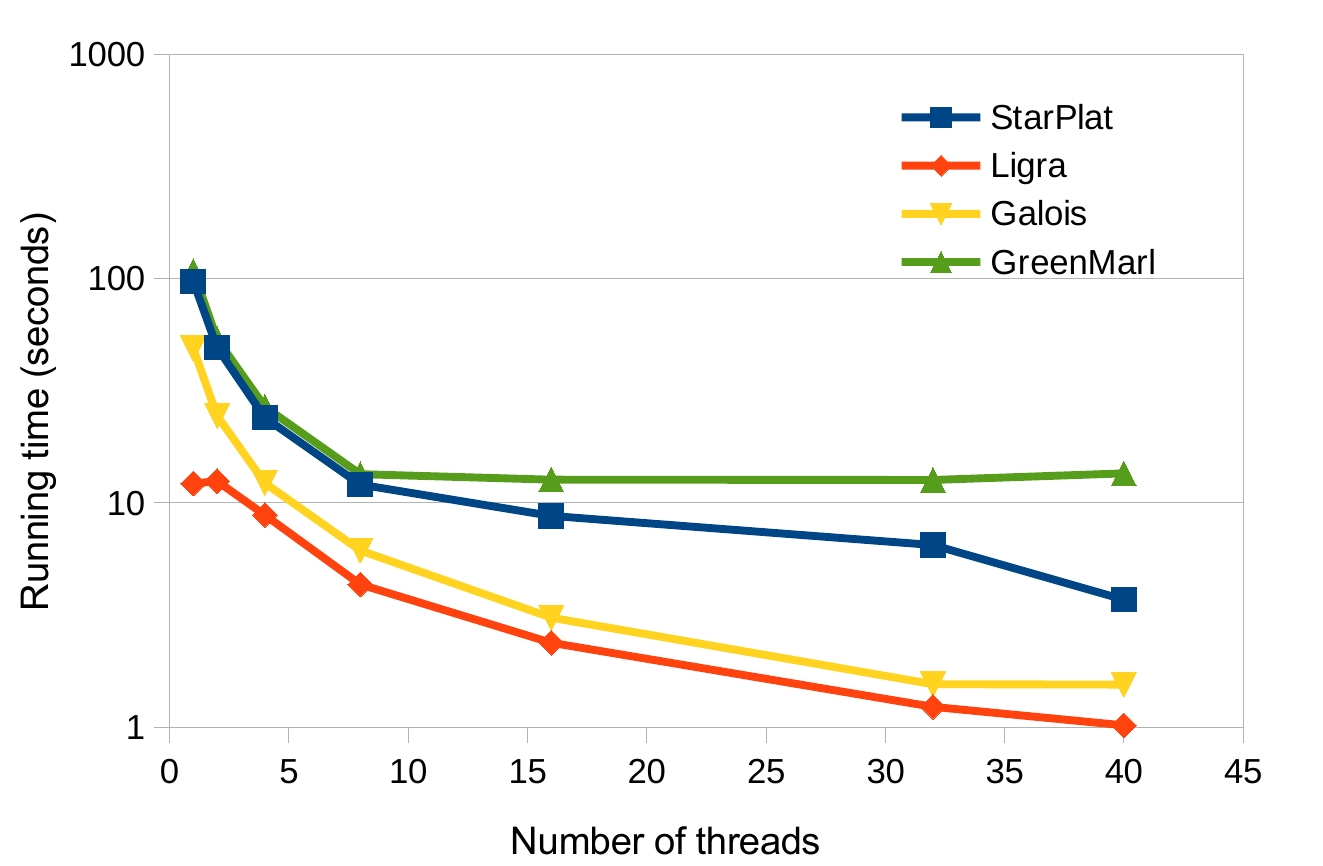}
    \caption{PK}
    \end{subfigure}
     \begin{subfigure}[b]{0.49\textwidth}
    \centering
    \includegraphics[scale=0.5]{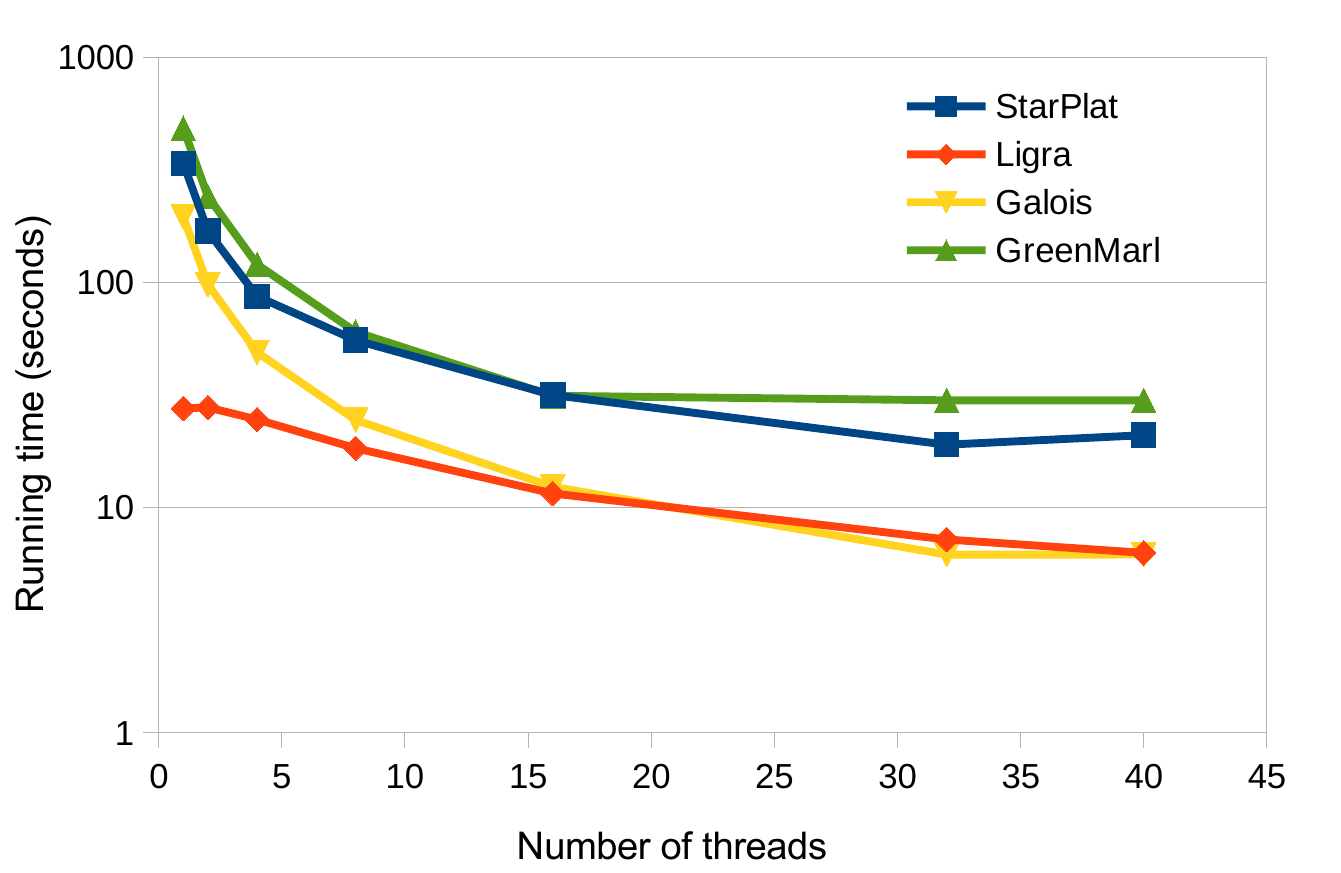}
    \caption{LJ}
    \end{subfigure}
    \caption{Log plots of running times of various types of graphs with varying number of threads (note different y-axis scales).
    }
    \label{fig:my_label}
\end{figure}

\subsection{MPI} \label{subsec exp eval mpi}

The comparison of execution time of \name's MPI code against the distributed Galois framework for four algorithms is given in Table~\ref{table:mpi-table}. The execution time does not include the graph reading and distribution time. OpenMPI 3.1.6 is used for execution and the algorithms are executed with 96 number of processes.












\arrayrulecolor{gray}
\begin{table}
\scalebox{0.93}{
\setlength{\tabcolsep}{3pt}
\begin{tabular}{|ll|l||r|r|r|r|r|r|r|r|r|r||r|}
\hline
\multicolumn{2}{|l|}{Algo.}                                      & Framework & \multicolumn{1}{c|}{TW} & \multicolumn{1}{c|}{SW} & \multicolumn{1}{c|}{OK} & \multicolumn{1}{c|}{WK} & \multicolumn{1}{c|}{LJ} & \multicolumn{1}{c|}{PK} & \multicolumn{1}{c|}{US} & \multicolumn{1}{c|}{GR} & \multicolumn{1}{c|}{RM} & \multicolumn{1}{c||}{UR} & \multicolumn{1}{c|}{Total} \\ \hline
\multicolumn{1}{|l|}{\multirow{6}{*}{BC}} & \multirow{2}{*}{20}  & Galois    & 2814                & 3502                & 1733                & 54776               & 2962                & 2331                & \textgreater 95 Hrs     & \textgreater 12 Hrs     & 1006                & 2611                & \multicolumn{1}{c|}{-}   \\
\multicolumn{1}{|l|}{}                    &                      & StarPlat  & \textbf{442}        & \textbf{13}          & \textbf{681}         & \textbf{244}         & \textbf{197}         & \textbf{52}         & \textbf{917}         & \textbf{353}        & \textbf{437}         & \textbf{75}         & 3411                 \\ \cline{2-14} 
\multicolumn{1}{|l|}{}                    & \multirow{2}{*}{80}  & Galois    & 12375               & 14043               & 6999                & \textgreater 12 Hrs     & 11320               & 9035                & \textgreater 12 Hrs     & \textgreater 12 Hrs     & 2479                & 10249               & \multicolumn{1}{c|}{-}   \\
\multicolumn{1}{|l|}{}                    &                      & StarPlat  & \textbf{2102}       & \textbf{49}         & \textbf{2705}       & \textbf{978}        & \textbf{764}         & \textbf{194}        & \textbf{3741}       & \textbf{1489}       & \textbf{975}         & \textbf{299}        & 13298                \\ \cline{2-14} 
\multicolumn{1}{|l|}{}                    & \multirow{2}{*}{150} & Galois    & NA & NA & NA & NA & NA & NA & NA & NA & NA & NA & -   \\
\multicolumn{1}{|l|}{}                    &                      & StarPlat  & \textbf{3816}         & \textbf{94}         & \textbf{4973}       & \textbf{1829}       & \textbf{1498}        & \textbf{382}        & \textbf{6782}       & \textbf{2713}       & \textbf{1442}        & \textbf{559}        & 24088                \\ \hline
\multicolumn{2}{|l|}{\multirow{2}{*}{PR}}                        & Galois    & 398                 & 761                 & 314                  & 376                  & 551                 & 479                 & 257                 & 296                 & 244                 & 268                 & 3943                 \\
\multicolumn{2}{|l|}{}                                           & StarPlat  & \textbf{57}         & OOM              & \textbf{23}         & \textbf{11}         & \textbf{8}          & \textbf{2}          & \textbf{2}          & \textbf{0.3}          & \textbf{44}         & \textbf{3}          & \multicolumn{1}{c|}{-}   \\ \hline
\multicolumn{2}{|l|}{\multirow{2}{*}{SSSP}}                      & Galois    & \textbf{33}         & 11                  & \textbf{30}         & 63                  & \textbf{27}         & 24                  & \textbf{1933}       & 82                  & \textbf{20}         & 49                  & 2272                 \\
\multicolumn{2}{|l|}{}                                           & StarPlat  & 152                 & \textbf{0.4}          & 104                 & \textbf{35}         & 47                  & \textbf{15}         & 2302                & \textbf{27}         & 70                  & \textbf{17}          & 2769                 \\ \hline
\multicolumn{2}{|l|}{\multirow{2}{*}{TC}}                        & Galois    & \textbf{71275}      & \textbf{436}        & \textbf{327}        & \textbf{884}        & \textbf{174}        & \textbf{19}          & 7                   & 9                   & \textbf{1238}       & \textbf{10}         & 74380                \\
\multicolumn{2}{|l|}{}                                           & StarPlat  & \textgreater 24 Hrs     & \textgreater 72 Hrs     & 64114                & \textgreater 24 Hrs     & 5741                & 2420                & \textbf{4}          & \textbf{0.6}          & \textgreater 24 Hrs     & 33                  & \multicolumn{1}{c|}{-}   \\ \hline
\end{tabular}
}
\caption{\name's MPI code performance comparison against Galois (96 processes). All times are in seconds. BC is run with the number of sources as \{20, 80, 150\}. \name goes out-of-memory on SW in PR due to double buffering.}
\label {table:mpi-table}
\end{table}

\mypara{\bcc} From the table, we can see that, \name's BC code takes much lesser time for execution compared to the Galois framework for all the ten graphs. When 20 number of sources are used, the execution of Galois code for two graphs TW and GR is not completed even after 96 hours and 12 hours respectively. Similarly, for 80 number of sources, the Galois code does not give result for three graphs WK, US and GR till 12 hours of execution. The execution of Galois code for all the 10 graphs with 150 number of sources is not tried as it is expected to take much more time than \name's BC code for all the ten graphs based on the execution time for 20 and 80 number of sources. 

\mypara{\prr} \name-generated PR code outperforms the Galois PR code for all the graphs except SW. \name's PR on SW goes out of memory due to double-buffering. We are currently investigating a way to fix it. 

\mypara{\ssspp}  From the comparison of execution time of SSSP code of \name with Galois, we can observe that, Galois performs better for 5 graphs TW, OK, LJ, US and RM while \name performs better for the remaining five graphs. The number of iterations taken by \name is fewer than that by Galois. This is due to multiple levels of distance update within the local nodes of a process in a single outer loop iteration.

\mypara{\tcc} The TC code generated by \name is less efficient as it iterates through neighbours of each node multiple times and check for neighbourhood among these nodes. Due to this, \name's TC code takes much more time than Galois. For road graphs US and GR, \name's TC takes lesser time than Galois code. For some graphs such as TW, SW, WK and RM \name's code did not give result even after 24 hours.

\subsection{CUDA} \label{subsec exp eval cuda}
Table~\ref{table:cuda} presents the absolute running times of the four algorithms on our ten graphs for the three frameworks: LonestarGPU, Gunrock, and \name.
The running times include CPU-GPU data transfer.

\arrayrulecolor{gray}
\begin{table}
\scalebox{0.9}{
\setlength{\tabcolsep}{3pt}
    \begin{tabular}{|lr|c|r|r|r|r|r|r|r|r|r|r||r|}
\hline
\multicolumn{2}{|l|}{Algo.} & Framework & TW & SW & OK & WK & LJ & PK & US & GR & RM & UR & Total \\ \hline
\multicolumn{1}{|l|}{\multirow{6}{*}{BC}} & 1 & LonestarGPU & - & - & - & - & - & - & - & - & - & - & - \\
\multicolumn{1}{|l|}{} & 1 & Gunrock & 2.122 & 4.237 & 0.525 & 0.535 & 0.548 & 0.317 & \textbf{2.811} & \textbf{1.750} & 1.238 & 0.944 & 15.027 \\
\multicolumn{1}{|l|}{} & 1 & \name & \textbf{0.002} & \textbf{0.004} & \textbf{0.149} & \textbf{0.153} & \textbf{0.078} & \textbf{0.029} & 17.656 & 6.359 & \textbf{0.225} & \textbf{0.079} & 24.734 \\ \cline{2-14} 
\multicolumn{1}{|l|}{} & 20 & \multirow{3}{*}{\name} & 6.992 & 2.279 & 2.762 & 3.014 & 1.298 & 0.534 & 369.701 & 126.485 & 2.949 & 1.593 & 517.607 \\
\multicolumn{1}{|l|}{} & 80 &  & 28.179 & 9.332 & 11.331 & 12.050 & 4.886 & 1.907 & 1444.656 & 518.968 & 6.509 & 6.372 & 2044.189 \\
\multicolumn{1}{|l|}{} & 150 &  & 55.548 & MLE & 21.241 & 27.271 & 9.609 & 3.794 & 2636.453 & 978.758 & 9.912 & 11.957 & 3754.543 \\ \hline
\multicolumn{2}{|l|}{\multirow{3}{*}{PR}} & LonestarGPU & - & \textbf{0.240} & 0.363 & \textbf{0.104} & \textbf{0.225} & \textbf{0.240} & \textbf{0.832} & \textbf{0.294} & \textbf{0.240} & \textbf{0.240} & 2.778 \\
\multicolumn{2}{|l|}{} & Gunrock & 15.230 & 36.910 & 2.430 & 2.460 & 2.952 & 1.085 & 13.345 & 6.499 & 9.170 & 5.487 & 95.568 \\
\multicolumn{2}{|l|}{} & \name & \textbf{4.081} & 7.112 & \textbf{0.256} & 1.780 & 1.300 & 0.257 & 3.420 & 0.679 & 0.891 & 0.257 & 20.033 \\ \hline
\multicolumn{2}{|l|}{\multirow{3}{*}{SSSP}} & LonestarGPU & 0.045 & 0.077 & 0.217 & 0.058 & 0.084 & 0.037 & \textbf{0.162} & \textbf{0.091} & 0.129 & 0.183 & 1.083 \\
\multicolumn{2}{|l|}{} & Gunrock & 2.272 & 4.057 & 0.616 & 0.556 & 0.562 & 0.311 & 1.283 & 1.140 & 1.034 & 0.915 & 12.746 \\
\multicolumn{2}{|l|}{} & \name & \textbf{0.001} & \textbf{0.002} & \textbf{0.078} & \textbf{0.044} & \textbf{0.027} & \textbf{0.012} & 1.667 & 0.695 & \textbf{0.120} & \textbf{0.028} & 2.674 \\ \hline
\multicolumn{2}{|l|}{\multirow{3}{*}{TC}} & LonestarGPU & - & 31.990 & 2.998 & 2.771 & \textbf{0.110} & \textbf{0.039} & 11.874 & 5.695 & \textbf{1.270} & 0.499 & 57.246 \\
\multicolumn{2}{|l|}{} & Gunrock & \textbf{67.718} & 7.369 & \textbf{0.843} & \textbf{0.997} & 0.850 & 0.404 & 1.490 & 0.712 & 3.200 & 1.040 & 84.623 \\
\multicolumn{2}{|l|}{} & \name & 10540.002 & \textbf{1.410} & 46.700 & 4.009 & 3.006 & 0.655 & \textbf{0.001} & \textbf{0.001} & 824.620 & \textbf{0.034} & 11420.430 \\ \hline
\end{tabular}
}
\caption{\name's CUDA code performance comparison against LonestarGPU and Gunrock. All times are in seconds. LonestarGPU does not have BC implemented and fails to load the largest graph TW.
}
\label {table:cuda}
\end{table}

\mypara{\bcc} \name-generated code outperforms Gunrock's library-based code on eight out of ten graphs. On the two road networks, Gunrock is considerably better than \name. 
We use CUDA's \texttt{grid.synchronization} to our benefit, whereas Gunrock relies heavily on this bulk-synchronous processing. Gunrock's Dijkstra's algorithm works very well for road networks. On the other hand, on social and random graphs, our implementation fares better.
We also illustrate performance with multiple sources of different sizes (20, 80, and 150). Except for soc-sinaweibo, \name-generated code is on par with or better than the other frameworks. 
Finally, unlike Gunrock and LonstarGPU, \name has the provision to execute BC from a set of sources.

\mypara{\prr}
We observe that the three frameworks have consistent relative performance, with hand-crafted LonestarGPU codes outperforming the other two and \name outperforming Gunrock. \name exploits the double buffering approach to read the current PR values and generate those for the next iteration. This separation reduces synchronization requirement during the update, but necessitates a barrier across iterations.
LonestarGPU uses an in-place update of the PR values and converges faster.

\mypara{\ssspp}
Gunrock uses Dijkstra's algorithm 
for computing the shortest paths using a two-level priority queue. We have coded a variant of the Bellman-Ford algorithm in \name. Hence, the comparison may not be most appropriate. But we compare the two only from the application perspective -- computing the shortest paths from a source in the least amount of time. LonestarGPU and \name outperform Gunrock on all the ten graphs. Between LonestarGPU and \name, there is no clear winner. They, in fact, have quite similar execution times.

\mypara{\tcc}
Unlike other three algorithms, TC is not a propagation based algorithm. In addition, it is characterized by a doubly-nested loop inside the kernel (see Figure~\ref{tc-dsl}). Another iteration is required for checking edge (Line~\ref{tc:isanedge}) which can be implemented linearly or using binary search if the neighbors are sorted in the CSR representation. Due this variation in the innermost loop, the time difference across various implementations can be pronounced, which we observe across the three frameworks. Their performances are mixed across the ten graphs, and no one emerges as an overall winner.

%% file: 6REL.tex
\section{Related Work} \label{sec relatedwork}
We divide the relevant related work based on the target backend in the next three subsections, and discuss generic frameworks in Section~\ref{sec:generic}.
.
\subsection{OpenMP} \label{subsec rel works omp}
Ligra~\cite{Ligra} is a lightweight graph processing framework that
is specific for shared-memory parallel/multi-core machines, which makes graph traversal
algorithms easy to write. The framework has two very simple routines, one for mapping over edges and one for mapping over vertices. The routines can be applied to any subset of the vertices, which makes the framework promising for many graph traversal algorithms operating on subsets of the vertices. In abstraction, a vertex subset is maintained as a set of integer labels for the included vertices, and the routine for mapping over vertices applies the user-supplied
function to each integer.

A lightweight infrastructure for graph analytics, Galois~\cite{Galois} supports a more general programming model which is more expressive than restrictive. The infrastructure supports autonomous scheduling of fine-grained tasks with application-specific priorities and a library for scalable data structures. Existing DSLs can be implemented on top of the Galois system to achieve high throughput. One can think of a Galois backend for \name.

Green-Marl~\cite{GreenMarl} is a DSL for shared memory graph processing. Programs depending on BFS/DFS traversals can be written concisely with the corresponding constructs in the language. However, for other graph algorithms, the user has to define the iteration over vertices or edges explicitly. So,  the algorithmic description can be written intuitively using the constructs provided by the language, while exposing the inherent data-level parallelism. \name borrows a few ideas from the language, and also quantitatively compares its performance against Green-Marl, Ligra, and Galois.

GraphIt~\cite{GraphIt} is a DSL for graph analysis that achieves high performance by enabling
programmers to easily find the best combination of optimizations for their specific algorithm
and input graph. It essentially separates computation from scheduling. The algorithms are specified using an algorithm language, and performance optimizations are specified using a separate scheduling language.

\subsection{MPI} \label{subsec rel works mpi}
D-Galois, a distributed graph processing system, was formed by interfacing Gluon, a communication optimization technique with Galois shared memory graph processing system~\cite{Gluon-DGalois}. We compare against Gluon in our experiments. 

Pregel and Giraph graph processing frameworks were inspired by the BSP programming model~\cite{Pregel,GiraphPaper}. 
Pregel decomposes the graph processing computation as a sequence of iterations in which the vertices receive data sent in the previous iteration, changes its state, and sends messages to its out neighbors which will be received in the next iteration. GPS provided additional features such as enabling global computation, dynamic graph repartitioning, and repartitioning of large adjacency lists to the Pregel framework~\cite{GPS}. 

A compiler automatically generates Pregel code from a subset of programs called Pregel canonical written using Green-Marl DSL~\cite{Pregel_GreenMarl}. It also applies certain transformations such as edge flipping, dissecting nested loops, translating random access, transforming BFS and DFS traversals to convert non-Pregel canonical pattern into Pregel canonical. Another compiler named DisGCo can translate all the Green-Marl programs to equivalent MPI RMA programs~\cite{DisGCo}. It handled the challenges such as syntactic differences, differences in memory view, intermixed serial code with parallel code, and graph representation while translating a Green-Marl program to MPI RMA code. 

The specification and implementation of morph graph algorithms and non-vertex centric graph algorithms for heterogeneous distributed environments was provided by the DH-Falcon DSL and its compiler~\cite{DH-Falcon}. A hybrid approach that uses both dense and sparse mode processing by utilizing pull- and push-based computation approaches was provided by a distributed graph analytics scheme Gemini~\cite{GeminiGraph}. Additionally, it used optimized graph representation and partitioning methods for intra-node as well as inter-node load balancing.

GraphLab provides a fault-tolerant distributed graph processing abstraction that can perform a dynamic, asynchronous graph processing for machine learning algorithms~\cite{GraphLab}. Another parallel distributed graph processing abstraction for processing power law graphs with a caching mechanism and a distributed graph placement mechanism was introduced by PowerGraph~\cite{PowerGraph}.
GraphChi supports asynchronous processing of dynamic graphs using a sliding approach on the disk-based systems~\cite{GraphChi}. 

\subsection{CUDA} \label{subsec rel works cuda}
Gunrock~\cite{Gunrock} is a graph library which uses data-centric abstractions to perform operations on edge and vertex frontier. 
 All Gunrock operations are bulk-synchronous, and they affect the frontier by computing on values within it or by computing a new one, using the following three functions: \textit{filter}, \textit{compute}, and \textit{advance}.  Gunrock library constructs efficient implementations of frontier operations with coalesced accesses and minimal thread divergence.

LonestarGPU~\cite{lonestargpu} is a collection of graph analytic CUDA programs. It employs multiple techniques related to computation, memory, and synchronization to improve performance of the underlying graph algorithms. We quantitatively compare \name against Gunrock and LonestarGPU.

Medusa~\cite{medusa} is a software framework which eases the work of GPU computation tasks. Similar to Gunrock, it provides APIs to build upon, to construct various graph algorithms. 
Medusa  exploits the BSP  model,  and  proposes a  new  model  EVM  (Edge  Message  Vertex), wherein  
the local  computations  are  performed  on  the  vertices and the computation progresses by passing messages across edges.  

CuSha~\cite{CUSHA} is a graph processing framework that uses two graph representations: G-Shards and Concatenated  Windows(CW). G-Shards makes use of a recently developed idea for non-GPU systems that divides a graph into ordered sets of edges known as shards. In order to increase GPU utilisation for processing sparse graphs, CW is a novel format that improves the use of shards. CuSha makes the most of the GPU's processing capability by processing several shards in parallel on the streaming multiprocessors. CuSha's architecture for parallel processing of large graphs allows the user to create the vertex-centric computation and plug it in, making programming easier.
CuSha significantly outperforms the state-of-the-art virtual warp-centric approach in terms of speedup.

MapGraph~\cite{MapGraph} is a parallel graph programming framework which provides a high level abstraction which helps in writing efficient graph programs. It uses SOA(Structure Of Arrays) to ensure coalesced memory access. It uses the dynamic scheduling strategy using GAS(Gather-Apply-Scatter) abstraction. Dynamic scheduling improves the memory performance and dispense the workload to the threads in accordance with degree of vertices.

\subsection{Generic Frameworks}\label{sec:generic}
Kakkos 3~\cite{kakkos-TrottLACDEGHHIL22} is a programming model in C++ for writing performance portable applications targeting all major HPC platforms~(CUDA, HIP, SYCL, HPX, OpenMP and C++ threads as backend). It provides abstractions for both parallel executions of code and data management. However, Kokkos kernels are used mainly on matrix-based operations or algorithms, and it targets coloring algorithms only recently, and many other complex algorithms are not supported (as of date). Philosophy-wise, Kokkos and \name have similarity, while the former is generic and is not a DSL. 

Graphit~\cite{GraphIt} adds  support for code generation on GPUs along with CPUs in their recent version. It expands the optimization space of GPU graph processing frameworks with a novel GPU scheduling language, and a compiler which enables various optimizations such as combining load balancing, edge traversal direction, active vertex set creation, active vertex set processing ordering, and kernel fusion. 

Julia~\cite{Julia-BezansonEKS17} is a high-level, dynamic programming language designed for high performance. It is well suited for numerical analysis and computational science. It supports concurrent, parallel, and distributed computing (with and without MPI). 
Circle~\cite{Circle-sean} is a new C++ 20 compiler. It extends C++ by adding many novel language features. Circle is a heterogeneous compiler that supports GPU compute and shader programming as well. With Circle, one can write a device-portable GPGPU code with Vulkan compute (using real pointers into physical storage buffers), a powerful feature only implemented by the Circle compiler. 
Carbon~\cite{Carbon-cpp} is an experimental successor language of C++ in development. One of its main goals is to support modern OS platforms, hardware architectures, environments, and fast \& scalable development (similar to Python and Rust). 
Odin~\cite{odin-lang} is again a general-purpose programming language with distinct typing built especially for high-performance, modern systems and data-oriented programming.  It is under development and supports structure of arrays~(SoA) data types and array programming. 
Chapel~\cite{Cheppel-praise,chapel-recent-HelbecqueG0MB22} is a modern parallel programming language aimed at portability and scalability. The primary goal of Chapel is to support general parallel programming and also make parallel programming at a scale far more productive. 

The recent sprouts of the above languages motivate the need for a simple language that is portable across different parallel programming platforms and architectures without compromising the productivity of the users, irrespective of the domain they work for.  

In a similar spirit, Intel~\cite{oneapi} and Nvidia~\cite{nvcpp} are venturing into heterogeneous programs which converts simple C++ program to parallel programs running on various architectures (multi-cores, GPUs, FPGAs and even Arm Servers). However, such programs are parallelized only if the program uses STL algorithms. Their compiler replaces the standard C++'s STL algorithms with the corresponding parallel versions based on the target.  Intel OneAPI~(which includes DPC++ and TBB) uses C++ STL, Parallel STL (PSTL), Boost Compute, and SYCL for parallelization. Nvidia's \texttt{nvc++} uses a different linker for the target processors using command-line arguments. \texttt{nvc++} supports ISO C++17, and targets GPU and multicore CPU programming with C++17 parallel algorithms, OpenACC, and OpenMP. 

We observe that Gunrock's Essentials-cpp and Essentials~\cite{Essentials-Osama:2022} are along similar lines and use modern C++, \texttt{std} parallelism constructs and lambdas built with their bulk-synchronous-asynchronous, data-centric abstraction model. They currently implement BFS and SSSP.


%% file: 7CON.tex
\section{Conclusion} \label{sec conclusion}
We presented \name, a DSL for implementing graph analytic algorithms. The language model provides various constructs for expressing graph problems at a high level without the complexities of implementing the commonly occurring patterns from scratch. This also facilitates providing optimized solutions both in terms of memory and time for these patterns. Currently, we support development across multicore, distributed, and manycore backends. We discussed the translation of \name's constructs for various backends. With a range of large graphs and four different algorithms, we experimentally validated that the performance of the generated code is comparable to the hand-tuned and generated implementations of existing frameworks and libraries. 

While there are multiple software solutions available in the form of programming languages and libraries for the development of graph algorithms, an important next step is to build solutions that are versatile to support multi-platform development. \name is an effort in this direction. 
The future direction of \name is to improve the portability of the solution to support OpenACC and OpenCL backends, and to enable automatic code generation for heterogeneous architectures.

%% file: 0-main-tpds.bbl

%% file: 91codes.tex
\section{Graph Algorithms in \name}\label{starplat:codes}
\name specifications for \bcc, \prr, and \tcc are shown below in Figures~\ref{bc-dsl}, \ref{pr-dsl}, and \ref{tc-dsl} respectively.
In addition, Figure~\ref{sssp-pull-dsl} shows pull-based SSSP.

\begin{lstlisting}[language=NEAR, style=mystyle, label=bc-dsl, caption = BC computation in \name , xleftmargin=.1\textwidth
]
function computeBC(Graph g, propNode<float> BC, SetN<g> sourceSet) {
    g.attachNodeProperty(BC = 0);
    for (src in sourceSet) {
        propNode<double> sigma;
        propNode<float> delta;
        g.attachNodeProperty(delta = 0);
        g.attachNodeProperty(sigma = 0);
        src.sigma = 1;

        iterateInBFS(v in g.nodes() from src) {
            for (w in g.neighbors(v)) {
                w.sigma = w.sigma + v.sigma;
        }   }

        iterateInReverse(v != src) {
            for (w in g.neighbors(v)) {
                v.delta = v.delta + (v.sigma / w.sigma) * (1 + w.delta);
            }
            v.BC = v.BC + v.delta;
}   }   }
\end{lstlisting}

\begin{lstlisting}[language=NEAR, style=mystyle, label=pr-dsl, caption = PR computation in \name , xleftmargin=.1\textwidth
]
function computePR(Graph g, float beta, float delta, int maxIter, propNode<float> pageRank) {
    float num_nodes = g.num_nodes();
    propNode<float> pageRank_nxt;
    g.attachNodeProperty(pageRank = 1 / num_nodes);
    int iterCount = 0;
    float diff;

    do {
        diff = 0.0;

        forall(v in g.nodes()) {
            float sum = 0.0;

            for (nbr in g.nodes_to(v)) {
                sum = sum + nbr.pageRank / g.count_outNbrs(nbr);
            }

            float val = (1 - delta) / num_nodes + delta * sum;
            diff += val - v.pageRank;
            v.pageRank_nxt = val;
        }

        pageRank = pageRank_nxt;
        iterCount++;

    } while ((diff > beta) && (iterCount < maxIter));
}
\end{lstlisting}

\begin{lstlisting}[language=NEAR, style=mystyle, label=tc-dsl, caption = TC computation in \name , xleftmargin=.1\textwidth
]
function computeTC(Graph g) {
    int triangle_count = 0;

    forall(v in g.nodes()) {
        forall(u in g.neighbors(v).filter(u < v)) {
            forall(w in g.neighbors(v).filter(w > v)) {
                if (g.is_an_edge(u, w)) { /*@\label{tc:isanedge}*/
                    triangle_count += 1;
}   }   }   }   }
\end{lstlisting}

\begin{lstlisting}[language=NEAR, style=mystyle, label=sssp-pull-dsl, caption = SSSP-Pull computation in \name , xleftmargin=.1\textwidth
]
function computePullSSSP (Graph g, node src) {
    propNode<int> dist;   
    propNode<bool> modified;  
    g.attachNodeProperty(dist = INF, modified = False);  
    src.modified = True;  
    src.dist=0;  
    bool finished = False;  
    fixedPoint until (finished: !modified) {   
        forall (v in g.nodes())  { 
            forall (nbr in g.nodes_to(v).filter(modified == True) {         
                edge e = g.getEdge(v, nbr); 
                <v.dist, v.modified> = <Min(v.dist, nbr.dist + e.weight), True>;  
}   }   }   }
\end{lstlisting}
